\RequirePackage{fixltx2e}
\RequirePackage{fix-cm}

\documentclass[%
floatfix,
showkeys,
nofootinbib, %
superscriptaddress, %
]{revtex4-1}

\usepackage{cmap}

\usepackage{ucs}
\usepackage[utf8x]{inputenc}
\usepackage[T1,T2A]{fontenc}
\usepackage[german,russian,english]{babel}

\usepackage[sort&compress]{natbib}
\usepackage{amsmath}
\usepackage{amssymb}

\usepackage{mathtools}
\mathtoolsset{
showonlyrefs,
mathic = true
}

\allowdisplaybreaks

\usepackage{hyperref}
\hypersetup{backref,
 colorlinks=false}
\hypersetup{pdfborder=0 0 0}

\usepackage{breakurl}

\usepackage{microtype}
\UseMicrotypeSet[protrusion]{alltext}

\usepackage{fixltx2e}

\usepackage{nicefrac}

\makeatletter
\def\ps@pprintTitle{%
     \let\@oddhead\@empty
     \let\@evenhead\@empty
     \let\@oddfoot\@empty
     \let\@evenfoot\@oddfoot}
\makeatother

\usepackage{amsthm} 
\usepackage{amsopn} 

\theoremstyle{definition}
\newtheorem*{thedefinition*}{Определение}

\theoremstyle{plain}
\newtheorem*{theorem*}{Теорема}

\newtheorem*{statement*}{Утверждение}

\newtheorem*{lemma*}{Лемма}

\theoremstyle{remark} 
\newtheorem*{theremark*}{Замечание}

\usepackage{mathrsfs}

\usepackage{comment}
\usepackage{spverbatim}

\usepackage{listings}
\usepackage{float}

\newcommand{\setR}{\mathbb{R}}
\renewcommand{\d}{\mathrm{d}}
\newcommand{\const}{\mathrm{const}} 
\begin{document}
\graphicspath{{image/rk-stoch-realis/en/}}

\title{Stochastic Runge--Kutta Software Package for Stochastic Differential Equations}

\author{M. N. Gevorkyan}
\email{mngevorkyan@sci.pfu.edu.ru} 
\affiliation{Department of Applied Probability and Informatics\\
Peoples' Friendship University of Russia\\
Miklukho-Maklaya str. 6, Moscow, 117198, Russia}

\author{T. R. Velieva}
\email{trvelieva@gmail.com}
\affiliation{Department of Applied Probability and Informatics\\
Peoples' Friendship University of Russia\\
Miklukho-Maklaya str. 6, Moscow, 117198, Russia}

\author{A. V. Korolkova}
\email{avkorolkova@gmail.com} 
\affiliation{Department of Applied Probability and Informatics\\
Peoples' Friendship University of Russia\\
Miklukho-Maklaya str. 6, Moscow, 117198, Russia}

\author{D. S. Kulyabov}
\email{yamadharma@gmail.com}
\affiliation{Department of Applied Probability and Informatics\\
Peoples' Friendship University of Russia\\
Miklukho-Maklaya str. 6, Moscow, 117198, Russia}
\affiliation{Laboratory of Information Technologies\\
Joint Institute for Nuclear Research\\
Joliot-Curie 6, Dubna, Moscow region, 141980, Russia}

\author{L. A. Sevastyanov}
\email{leonid.sevast@gmail.com}
\affiliation{Department of Applied Probability and Informatics\\
Peoples' Friendship University of Russia\\
Miklukho-Maklaya str. 6, Moscow, 117198, Russia}
\affiliation{Bogoliubov Laboratory of Theoretical Physics\\
Joint Institute for Nuclear Research\\
Joliot-Curie 6, Dubna, Moscow region, 141980, Russia}

\begin{abstract}

  As a result of the application of a technique of multistep processes stochastic models construction 
  the range of models, implemented as a self-consistent 
  differential equations, was obtained. These are partial differential equations (master equation, 
  the Fokker--Planck equation) and stochastic differential equations (Langevin equation). 
  However, analytical methods 
  do not always allow to research these equations adequately. It is proposed to use the combined 
  analytical and numerical approach studying  these equations. 
  For this purpose the numerical part is realized within the framework of symbolic computation. 
   It is recommended to apply stochastic Runge--Kutta methods for numerical study of stochastic 
   differential equations in the form of the Langevin. Under this approach, 
  a program complex on the basis of analytical calculations metasystem Sage is developed. 
  For model verification  logarithmic walks and Black--Scholes two-dimensional model are used. 
  To illustrate the stochastic ``predator--prey'' type model is used. 
  The utility of the combined numerical-analytical approach is demonstrated.
  \end{abstract}


\maketitle

\section{Introduction}

The mathematical models adequacy may be largely increased by taking into account 
stochastic properties of dynamic systems. Stochastic models are widely
used in chemical kinetics, hydrodynamics, population dynamics,
epidemiology, filtering of signals, economics and financial
mathematics as well as different fields of
physics~\cite{Kloeden_Platen}.  \emph{Stochastic differential
  equations} (SDE) are the main mathematical apparatus of such models.

Even for the case of ordinary differential equations (ODE) an analytical
solution can be obtained only for a limited class of equations. That's
why a great variety of numerical methods have been
developed~\cite{Butcher_2003,Hairer:1::en} for applications. In the case
of SDE the importance of numerical methods greatly increases, because
the exact analytical solutions have been obtained only for a very
small number of stochastic models~\cite{Kloeden_Platen,Andreas_2003}.

Compared with numerical methods for ODEs, 
numerical methods for SDEs are much less developed. There are
two main reasons for this: a comparative novelty of this field of
applied mathematics  and much more complicated mathematical
apparatus. The development of new numerical methods for stochastic case
in many ways is similar to deterministic methods development 
\cite{Burrage_1996,Burrage_1997,Burrage_1998,Burrage_2000,Rossler_2010,Andreas_2003}. The
scheme, that has been proposed by Butcher~\cite{Butcher_2003}, gives
visual representation of three main classes of numerical schemes.


The accuracy of numerical scheme may be improved in the following
way: by adding additional steps (multi-step), stages (multi-stage),
and the derivatives of drift vector and diffusion matrix
(multi-derivative) to the scheme.

Multi-step (Runge--Kutta like) numerical methods are more suitable for
the program implementation, because they can be expressed as a
sequence of explicit formulas. Thus, it is natural to spread
Runge--Kutta methods in the case of stochastic differential equations.

The main goal of this paper is to give a review of stochastic Runge--Kutta 
methods implementation made by authors for Sage computer algebra
system~\cite{sage}. The Python programming language and NumPy and SciPy 
modules were used for this purpose because Python programming language is an 
open and powerful framework for scientific calculations.

A few additional algorithms have to be implemented before proceeding to realisation of
stochastic Runge-Kutta algorithm. These algorithms are:
generation of Wiener process trajectories, approximation of 
multiple stochastic integrals, n-point distributed random values generation. 

The authors have faced the necessity of stochastic numerical
methods  implementation in the course of work on the method of
stochastization of one-step
processes~\cite{kulyabov:2013:conf:mmcp,kulyabov:2014:icumt-2014:p2p,ef-kor-gev-kul-sev:vestnik-miph:2014-3}. As
 The  ``predator-prey''-type  model was taken as an example of the methodology application in order to 
verify the obtained SDE models by numerical experiments.  The same problem had aroused 
when authors developed stochastic models for
RED-like queuing disciplines~\cite{kulyabov:2014:icumt-2014:gns3}.

\subsection{Article's content}

In the first part of the article a brief introduction of
SDE's mathematical apparatus is given. The definition of the scalar and
multidimensional Wiener process, the definition of the scalar and
multidimensional Itō SDEs, strong and weak convergence of
the approximating function, matrix formula for double Itō integrals
approximation are introduced.

The second part of the article is devoted to the description of
stochastic numerical Runge--Kutta  schemes with strong and weak orders
of convergence for scalar and multi-dimensional Wiener processes.

The third part provides necessary information about library functions for implementation of 
numerical schemes from the second part of the paper.

In the final section the introduced in previous parts of the article methods are used 
to derive strong and weak
approximations of ``predator-prey'' stochastic model.  Also some
conclusions about dissimilarity the stochastic version of ``predator-pray'' models from 
deterministic version are made based on numerical results.

The article can be regarded as a practical introduction to
the field of stochastic numerical methods. Also it  could be useful as
guide for the implementation of stochastic numerical schemes in any
other programming languages.

\subsection{The choice of programming language for implementing stochastic numerical methods}

The following requirements to computer algebra 
system's programming language were taken into account:
\begin{itemize}
\item advanced tools for manipulations with multidimensional arrays
  (up to four axes) with a large number of element are needed;
\item it is critical to be able to implement parallel execution of
  certain functions and sections of code due to the need of a large
  number of independent calculations according to Monte-Carlo method;
\item functions to generate large arrays of random numbers are needed;
\item software should be open and free for
   non-commercial use.
\end{itemize}

Computer algebra system Sage generally meets all these
requirements.  NumPy module is used for array manipulations and SciPy module --- 
for n-point distribution creation.

\section{Wiener stochastic process}

In this section only the most important definitions and
notations are introduced. For brief introduction to SDE
see~\cite{Higham_2001,Malham_2010}, for more details
see~\cite{Kloeden_Platen,Platen_Liberati_2010,Oksendal::en}.

A standard scalar Wiener process $W(t)$, $t \geqslant 0$ is a
stochastic process which depends continuously on $t\in[t_{0},T]$ and
satisfies the following three conditions:
\begin{itemize}
\item $\mathrm{P}\{W(0)=0\} = 1$, in other words $W(0) = 0$ almost surely;
\item $W(t)$ has independent increments i.e.
        $\{\Delta W_{i}\}^{N-1}_{0}$~--- independently distributed random variables;
        $\Delta W_{i} = W(t_{i+1}) - W(t_{i})$ and
        $0 \leqslant t_{0} < t_{1} < t_{2} < \ldots < t_{N} \leqslant T$;
\item
        $\Delta W_{i} = W(t_{i+1}) - W(t_{i}) \sim
        \mathcal{N}(0,t_{i+1}-t_{i})$,
        where $0\leqslant t_{i+1} < t_{i} < T$, $i=0,1,\ldots,N-1$.
\end{itemize}

Notation $\Delta W_{i} \sim \mathcal{N}(0,\Delta t_{i})$ denotes that
$\Delta W_{i}$ is a normally distributed random variable with zero
mean $\mathbb{E}[\Delta W_{i}] = \mu = 0$ and unit variance
$\mathbb{D}[\Delta W_{i}] = \sigma^{2} = \Delta t_{i}$.

Multidimensional Wiener process
$W^{\alpha}(t)\colon \Omega\times[t_{0},T]\to \mathbb{R}^{m},\;\forall
\alpha = 1,\ldots,m$,
is consist of independent $W^{1}(t),\ldots,W^{m}(t)$ scalar Wiener
processes. The increments $\Delta W^{\alpha}_{i}$ are mutually
independent normally distributed random variables with zero mean and
unit variance.

\subsection{Computer simulation of sample Wiener process paths}

\begin{figure}
  \centering
  \includegraphics[width=0.4\linewidth]{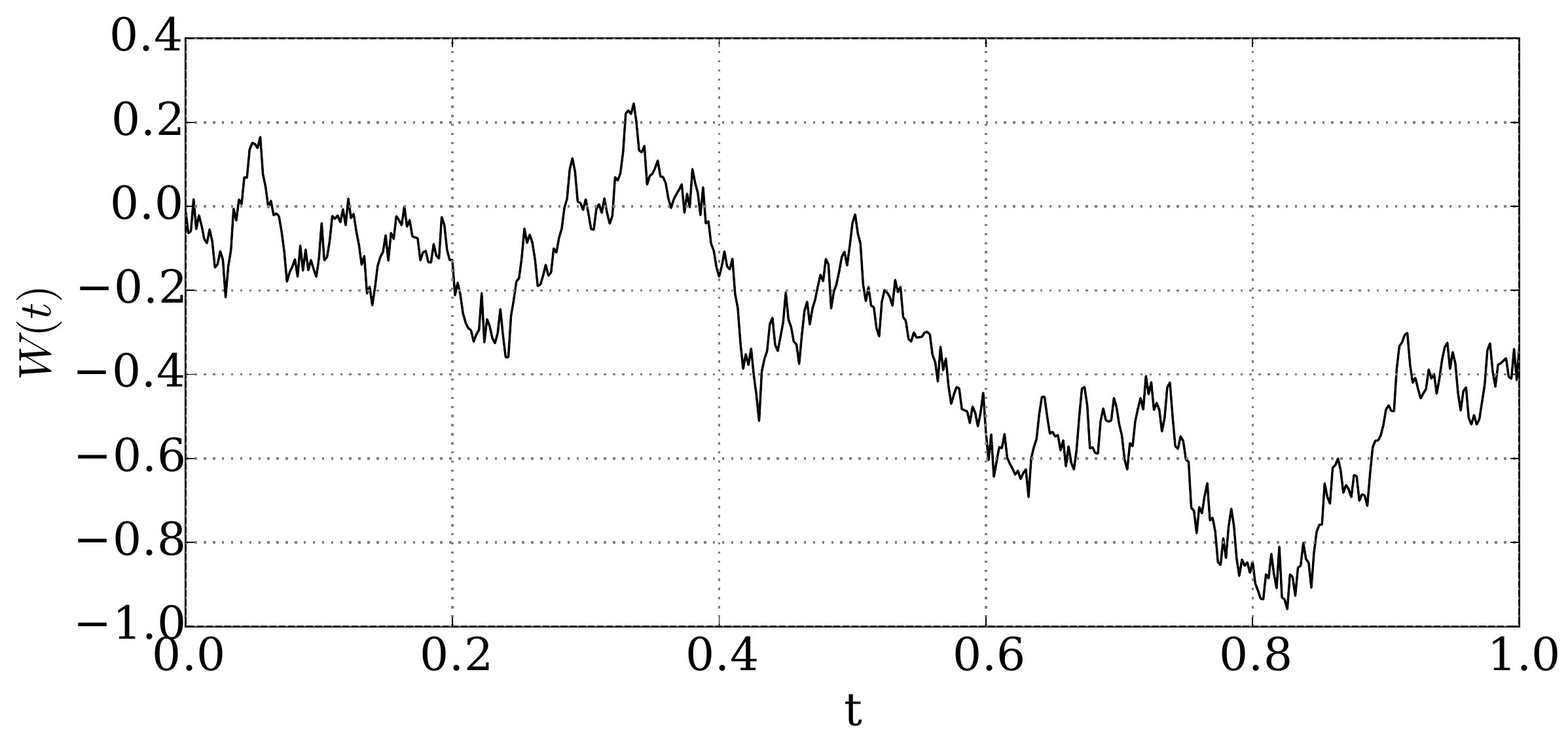}
  \caption{Wiener process path}
\label{fig:wiener}
\end{figure}

In order to construct Wiener process sample path on $[0,T]$ interval divided into $N$ parts with 
subinterval step $\Delta t_{i} = h = \mathrm{const}$, we need to generate $N$
normally distributed random numbers
$\varepsilon_{0}, \ldots, \varepsilon_{N-1} \sim \mathcal{N}(0,h)$ 
and find their cumulative sums $\varepsilon_{0}$,
$\varepsilon_{0} + \varepsilon_{1}$,
$\varepsilon_{0} + \varepsilon_{1} + \varepsilon_{2}$ and so on. As the result 
two arrays (each with $N$ elements) will be get. The first array \texttt{W} consist of 
Wiener sample path points and the second \texttt{dW} is an array of increments. 
For example of such trajectory see fig.~\ref{fig:wiener}.

In case of $m$-dimension process $m$ sequences of $N$ normally
distributed random variables should be generated.

В описываемой нами библиотеке для генерирования выборочных
винеровских траекторий написана функция:
\begin{lstlisting}[language=python]
(dt, t, dW, W) = wiener_process(N, dim=1, interval=(0.0, 1.0))
\end{lstlisting}
Обязательны аргумент $N$ --- число точек разбиения временного
интервала \texttt{interval} (по умолчанию $[0,1]$), а \texttt{dim}
--- размерность винеровского процесса. Функция возвращает кортеж из
четырех элементов, где \texttt{dt} --- шаг разбиения
($\Delta t_{i} = h = \mathrm{const}$), \texttt{t} --- одномерный
\texttt{numpy} массив содержащий значения моментов времени
$t_{1},t_{2},\ldots,t_{N}$, \texttt{dW}, \texttt{W} --- также
\texttt{numpy} массивы размерностью $N\times m$ содержащие
приращения $\Delta W^{\alpha}_{i}$ и точки выборочной траектории
$W^{\alpha}_{i}$, где $i=1,\ldots,N;\;\alpha=1,\ldots,m$.

\subsection{It\^{o} SDE for multidimensional Wiener process}

Let's consider a random process
$\mathbf{x}(t) = (x^{1}(t), \ldots, x^{d}(t))^{T}$, where
$\mathbf{x}(t)$ belongs to the function space
$\mathrm{L}^{2}(\Omega)$ with norm $\| \cdot \|$. We assume that
stochastic process $\mathbf{x}(t)$ is a solution of Itō SDE~\cite{Oksendal::en,Kloeden_Platen}:
\[
x^{\alpha}(t) = f^{\alpha}(t,x^{\gamma}(t))\d t + \sum\limits_{\beta =
  1}^{m}G^{\alpha}_{\beta}(t,x^{\gamma}(t))\d W^{\beta},
\]
where $\alpha,\gamma = 1,\ldots,d$, $\beta = 1,\ldots,m$, vector value function
$f^{\alpha}(t,x^{\gamma}(t)) =$ $f^{\alpha}(t,x^{1}(t),\ldots,x^{d}(t))$
is a \emph{drift vector}, and matrix value function
$g^{\alpha}_{\beta}(t,x^{\gamma}(t))$ is a \emph{diffusion matrix},
$W^{\alpha} = (W^{1},\dots,W^{m})^T$ is a multidimensional Wiener
process, also known as a \emph{driver} process of SDE.

Let's introduce the discretization of interval $[t_{0}, T]$ by sequence 
$t_{0}<t_{1}<\ldots<t_{N}=T$ with step $h_{n} = t_{n+1} - t_{n}$, where 
$n = 0,\ldots,N-1$ and $h = \max{\{h_{n-1}\}^{N}_{1}}$ --- minimal step. 
We also consider the step $h_{n} = h$ to be a constant; $\mathbf{x}_{n}$ ---
mesh function for random process $\mathbf{x}(t)$ approximation,
i.e. $\mathbf{x}_{0} = \mathbf{x}(t_{0})$,
$\mathbf{x}_{n} \approx \mathbf{x}(t_{n})\;\forall n = 1,\ldots,N$.

\subsection{Strong and weak convergences of approximation function}

It is necessary to define the criterion for measuring the accuracy of process
$\mathbf{x}(t)$  approximation by sequences of functions  
$\{\mathbf{x}_{n}\}^{N}_{1}$. Usually 
\emph{strong} and
\emph{weak}~\cite{Oksendal::en,Andreas_2003,Rossler_2010} criteria are
defined.

The sequence of approximating functions
$\{\mathbf{x}_{n}\}^{N}_{1}$ converges  with order
        $p$ to exact solution $\mathbf{x}(t)$ of SDE in moment $T$ in \emph{strong sense} if constant 
        $C>0$ exists and $\delta_{0}>0$, such as
$\forall h \in (0,\delta_{0}]$ the condition
\[
\mathbb{E}(\|\mathbf{x}(T) - \mathbf{x}_{N}\|) \leqslant Ch^{p}
\]
is fulfilled.

The sequence of approximating functions
$\{\mathbf{x}_{n}\}^{N}_{1}$ converges  with order
$p$ to solution $\mathbf{x}(t)$ of SDE in moment $T$ in \emph{weak seance} if constant 
                        $C_{F}>0$ exists and $\delta_{0}>0$, such as
$\forall h \in (0,\delta_{0}]$ the condition
\[
\left|\mathbb{E}\left[F(\mathbf{x}(T))\right] -
        \mathbb{E}\left[F(\mathbf{x}_{N})\right]\right| \leqslant
C_{F}h^{p}
\]
is fulfilled.

$F \in C^{2(p+1)}_{\mathrm{P}}(\setR,\setR^{d})$ is continuously 
differentiable (including $2(p+1)$-th derivatives) functional with
polynomial growth.

If the matrix $\mathbf{G}$ vanishes, the condition of strong
convergence is equivalent to the deterministic
case convergence condition. In contrast to the deterministic case the order of strong
convergence is not necessarily a natural number and may
take the rational fractional values.

The convergence type selection depends entirely on the problem being solved. 
Numerical methods with a strong convergence are the best choice
for specific trajectory approximation of a random process
$\mathrm{x}(t)$ and therefore the information on the driving Wiener
process ia necessary. In practice, this means that the function, that implements the
method with strong stochastic convergence, must get two-dimensional
array with increments of $\Delta W^{\alpha}_{n} $, where
$\alpha=1,\ldots,m;\;n=1,\ldots,N$.  This array will be also used for
the approximation of multiple stochastic integrals.

Weak numerical methods are suitable for approximation of characteristics of the 
random process $\mathrm{x}(t)$ probability
distribution, because no information about the trajectory driving Wiener process
is required, and
random values for these methods can be generated on some other
probability space, which is easy to implement in software.

\subsection{Approximation of It\^{o} multiple stochastic integrals}

In general, for the design of numerical schemes with the order of the strong
convergence grater than $p=\frac{1}{2}$, it is necessary to include
single and double It\^{o} integrals in the these schemes formula:
\begin{gather}
I^{i}(t_{n}, t_{n+1}) = I^{i}(h_{n}) = \int\limits_{t_{n}}^{t_{n+1}}
\mathrm{\d}W^{i}(\tau),\\
I^{ij}(t_{n}, t_{n+1}) = I^{ij}(h_{n}) 
=  
\int\limits_{t_{n}}^{t_{n+1}}\int\limits_{t_{n}}^{\tau_{1}}\mathrm{d}W^{i}(\tau_{2})\mathrm{d}W^{j}(\tau_{1}),
\end{gather}
where $i,j=1\ldots,m$ and $W^{i}$ are components of Wiener process.

So the main task is to express It\^{o} integrals in terms of
 Wiener process increments $\Delta W^{i}_{n} = W^{i}(t_{n+1}) - W^{i}(t_{n})$. 
 In the case of single integral it is possible for any index $i$:
$I^{i}(h_{n}) = \Delta W^{i}_{n}$. The case of double $I^{ij}(h_{n})$ integral 
is more complicated and there is an exact formula only for $i=j$ case:
\[
I^{ii}(h_{n}) = \dfrac{1}{2}\left( (\Delta W^{i}_{n})^2 - \Delta t_{n}
\right).
\]
In other cases, i.e. $i\neq j$ for $I^{ij}(h_{n})$, the exact expression with
increments $\Delta W^{\alpha}_{n}$ and $\Delta t_{n}$ do not exist, 
so the numerical approximation shell be used.

In paper~\cite{Wiktorsson_2001} author introduced a matrix form of approximation formulas. 
Let's denote as $\mathbf{1}_{m\times m}$ and $\mathbf{0}_{m\times m}$ the unit and zero 
matrices $m\times m$. Then
\begin{gather}
  \mathbf{I}(h_{n}) = \dfrac{\Delta \mathbf{W}_{n} \Delta
    \mathbf{W}_{n} - h_{n}\mathbf{1}_{m\times m}}{2} +
  \mathbf{A}(h_{n}), \\
  \mathbf{A}(h_{n}) =
  \dfrac{h}{2\pi}\sum\limits^{\infty}_{k=1}\dfrac{1}{k}\biggl(\mathbf{V}_{k}(\mathbf{U}_{k}
    + \sqrt{2/h_{n}} \Delta \mathbf{W}_{n})^{T} 
    - 
    (\mathbf{U}_{k} +
    \sqrt{2/h_{n}}\Delta \mathbf{W}_{n})\mathbf{V}^{T}_{k}\biggr),
\end{gather}
where $\Delta \mathbf{W}_{n}, \mathbf{V}_{k}, \mathbf{U}_{k}$ are normally 
distributed independent multidimensional random values:
\begin{gather}
  \Delta \mathbf{W}_{n} = (\Delta W^{1}_{n}, \Delta W^{2}_{n}, \ldots,
  \Delta W^{m}_{n})^{T} \sim \mathcal{N}(\mathbf{0}_{m\times m},
  h_{n}\mathbf{1}_{m\times m}),
  \\
  \mathbf{V}_{k} = (V^{1}_{k}, V^{2}_{k},\ldots, V^{m}_{k})^{T} \sim
  \mathcal{N}(\mathbf{0}_{m\times m}, \mathbf{1}_{m\times m}),
  \\
  \mathbf{U}_{k} = (U^{1}_{k}, U^{2}_{k},\ldots, U^{m}_{k})^{T} \sim
  \mathcal{N}(\mathbf{0}_{m\times m}, \mathbf{1}_{m\times m}).
\end{gather}

\subsection{SDEs with exact solutions}

For verification of written programs in the case of strong convergence 
SDEs with a known analytical solution will be used, so  it makes possible to compute the 
error of trajectories approximation.

        \subsection{Logarithmic walk (one-dimensional Wiener process)}

As an example of SDE with known exact solution the 
\emph{logarithmic walk} model~\cite{Platen_Liberati_2010,Kloeden_Platen} is used:
\[
\mathrm{d}x(t) = \mu x(t) \mathrm{d}t + \sigma x(t)
\mathrm{d}W(t),\;\; x(0) = x_{0},
\]
the exact solution is~\cite[section 4.4]{Kloeden_Platen}:
\[
x(t) = x_0 \exp{\left((\mu - 0.5\sigma^2)t + \sigma W(t)\right)}.
\]
Initial values are $\mu = 2,\; \sigma = 1$ and
$ x_{0} = 1$.

\subsection{Two-dimensional Black--Scholes model}

        For two dimensional Wiener process the two-dimensional 
                Black--Scholes model~\cite{Platen_Liberati_2010} is used in following form:
\[
\mathbf{x}(t) = \mathrm{A}\mathbf{x}(t)\mathrm{d}t +
\sum\limits^{2}_{k=1}\mathrm{B}_k \mathbf{x}(t)\mathrm{d}W^{k},
\]
all matrices are diagonal: $A = \mathrm{diag}(a^1,a^2)$,
$B_{1} = \mathrm{diag}(b^1_1,b^2_1)$ and
$B_{2} = \mathrm{diag}(b^1_2,b^2_2)$. Drift vector $\mathbf{f}(\mathbf{x})$ 
and diffusion matrix $\mathrm{G}(\mathbf{x})$ can be written as
\[
\mathrm{f}(\mathbf{x}) =
\begin{bmatrix}
  a^{1}x^{1}(t)\\
  a^{2}x^{2}(t)
\end{bmatrix},\;\; \mathrm{G}(\mathbf{x}) =
\begin{bmatrix}
  b^1_1 x^{1}(t)&b^1_2 x^{1}(t)\\
  b^2_1 x^{2}(t)&b^2_2 x^{2}(t)
\end{bmatrix}.
\]
The exact analytical solution is  given by
\[
\mathbf{x}(t) = \mathbf{x}_0 \exp \left\{\left(\mathrm{A} -
    \frac{1}{2}\sum\limits^{2}_{k=1}B_k^2\right)t +
  \sum\limits^{2}_{k=1}\mathrm{B}_k W^k (t)\right\}.
\]
For calculations we will take matrices:
$A = \mathrm{diag}(a_1,a_2)$, $B_{1} = \mathrm{diag}(b_1,b_2\rho)$, 
$B_{2} = \mathrm{diag}(0,b_2\sqrt{1-\rho})$, 
and following values of parameters $x^{1}_{0} = x^{1}_{0} = 1.0$,
$a_{1} = a_{2} = 0.1$, $b_{1} = b_{2} = 0.2$, $\rho = 0.8$.  
Then the exact solution of the equation takes the form:
\begin{align*}
  &x^{1}(t) = x^{1}_{0}\exp\left((a_{1} - 0.5b^{2}_{1})t + b_{1}W^{1}(t)\right)\\
  &x^{2}(t) = x^{2}_{0}\exp\left((a_{2} - 0.5b^{2}_{2})t + b_{2}\left(\rho W^{1}(t) + \sqrt{1+\rho^2}W^{2}(t)\right)\right).
\end{align*}

\section{Stochastic Runge-Kutta methods}

In this section some of the most effective
stochastic numerical Runge--Kutta methods are presented (based on the
papers~\cite{Rossler_2010,Andreas_2003,Debrabant_2007,Debrabant_2013,Tocino_2001,Soheili_2007,Mackevicius_1994}). We
restrict ourselves
 to the numerical schemes without derivatives of
drift vector and  diffusion matrix, so the effective Milstein schemes 
will be neglected~\cite{Milstein_1974,Milstein_1979,Milstein_1986}.

In the beginning we will mention some factors that make stochastic
Runge--Kutta methods more complicated in comparison to classical methods:
\begin{itemize}
\item When selecting a particular method we must take into account what type of
convergence is necessary to provide for this specific stochastic model, and
also which type of the stochastic equations is to be solved ---
It\^{o} form or the Stratonovich form. This increases the number of
algorithms to be programmed.
\item For methods with a strong convergence of $p_{s}>1$ the approximation of double
stochastic integrals is required and this is a very expensive computational task.
\item In the numerical scheme there are not only matrices and vectors but
also tensors (four-dimensional arrays). So convolution operation on several 
indices is required. Implementation of convolution operation as
summation by means of cycles leads to a significant drop of performance.
\item In order to use the weak methods, Monte Carlo method should be applied
and several series of calculations ($10^7$-$10^8$ calculations in each series) are required.
\end{itemize}

The program operation speed depends greatly on the necessity of multidimensional 
arrays convolution. \texttt{NumPy} 
module provides very useful function \texttt{einsum} (Einstein summation) 
which greatly improves the speed of convolution calculation.

It is also important to pay attention to the large number of zeros in
generalized Butcher tables. Implementation of a separate function for
specific method often helps to get a performance goal in comparison 
with the function that implements a universal algorithm.

\subsection{Euler--Maruyama numerical method}

The simplest numerical method for solving both scalar equations and 
systems of SDEs is Euler-Maruyama method, named in honor of
Gishiro Maruyama (Gisiro Maruyama), who extends the classical
Euler's method from ODE to SDE case~\cite{Maruyama_1955}. The method is easily
generalized to the case of multidimensional Wiener process:
\begin{align*}
&x^{\alpha}_{0} = x^{\alpha}(t_{0}),\\
&x^{\alpha}_{n+1} = x^{\alpha}_{n} + f^{\alpha}(t_{n},
x^{\alpha}_{n})h_{n} +
\sum\limits^{d}_{\gamma=1}G^{\alpha}_{\beta}(t_{n},x^{\gamma}_{n})\Delta
W^{\beta}_{n}.
\end{align*}

As seen from the formulas, on each step only the increment of 
the driving Wiener process $\Delta W^{\beta}_{n} $ is required.
The method has a strong $(p_ {d}, p_ {s}) = (1.0, 0.5)$ order. 
The $p_ {d}$ value  denotes the accuracy order of the deterministic 
numerical method, i.e. the precision that a numerical method will give being applied 
to the equation with $G(t, x^{\alpha}(t))\equiv 0$. The $p_{s}$ value 
denotes the order of approximation of the stochastic part of the equation.

\subsection{Strong stochastic Runge--Kutta methods for SDEs with
  scalar Wiener process}

In the case of scalar SDE and Wiener process the following scheme is valid:
\begin{align}
&X^{i}_{0} = x_{n} + \sum\limits^{s}_{j=1}A_{0j}^{i}f(t_{n} + c^{j}_{0}h_{n}, X^{j}_{0})h_{n} + \sum\limits^{s}_{j=1}B^{i}_{0j}g(t_{n} + c^{j}_{1}h_{n}, X^{j}_{1})\dfrac{I^{10}(h_{n})}{\sqrt{h_{n}}},\\
&X^{i}_{1} = x_{n} + \sum\limits^{s}_{j=1}A_{1j}^{i}f(t_{n} + c^{j}_{0}h_{n}, X^{j}_{0})h_{n} + \sum\limits^{s}_{j=1}B^{i}_{1j}g(t_{n} + c^{j}_{1}h_{n}, X^{j}_{1})\sqrt{h_{n}},\\
&x_{n+1} = x_{n} + \sum\limits^{s}_{i=1}a_{i}f(t_{n}+c^{i}h_{n}, X^{i}_{0})h_{n} + \\
& \qquad + \sum\limits^{s}_{i=1}\left(b^{1}_{i}I^{1}(h_{n}) + b^{2}_{i}\dfrac{I^{11}(h_{n})}{\sqrt{h_{n}}} + b^{3}_{i}\dfrac{I^{10}(h_{n})}{h_{n}} + b^{4}_{i}\dfrac{I^{111}(h_{n})}{h_{n}}\right)g(t_{n} + c^{i}_{1},X^{i}_{1}).
\end{align}
The method is characterized by the Butcher table~\cite{Rossler_2010}:
\[ {\renewcommand{\arraystretch}{1.5}%
\begin{array}{c|c|c|c}
c^{i}_{0} & A^{i}_{0j} & B^{i}_{0j} &\\ \hline
c^{i}_{1} & A^{i}_{1j} & B^{i}_{1j} &\\ \hline
                                        & a_{i} & b^{1}_{i} & b^{2}_{i}\\ \hline
                                        &  & b^{3}_{i} & b^{4}_{i}\\
\end{array}}.
\]

In Rößler preprint~\cite{Rossler_2010} there are two realisations of 
this scheme for $s=4$:
{\scriptsize%
\[
{\renewcommand{\arraystretch}{1.5}%
\begin{array}{c|cccc|cccc|cccc}
        0&0&0&0&0&0&0&0&0&&&\\
        \frac34&\frac34&0&0&0&\frac{3}{2}&0&0&0&&&\\
        0&0&0&0&0&0&0&0&0&&&\\
        0&0&0&0&0&0&0&0&0&&&\\ \hline
        0&0&0&0&0&0&0&0&0&&&\\
        \frac14&\frac14&0&0&0&\frac12&0&0&0&&&\\
        1&1&0&0&0&-1&0&0&0&&&\\
        \frac14&0&0&\frac14&0&-5&3&\frac12&0&&&\\ \hline
        &\frac13&\frac23&0&0&-1&\frac43&\frac23&0&-1&\frac43&-\frac13&0\\ \hline
        &&&&&2&-\frac43&-\frac23&0&-2&\frac53&-\frac23&1
\end{array}}\;\;
{\renewcommand{\arraystretch}{1.5}%
\begin{array}{c|cccc|cccc|cccc}
        0&0&0&0&0&0&0&0&0&&&\\
        1&\frac34&0&0&0&0&0&0&0&&&\\
        \frac12&\frac14&\frac14&0&0&1&\frac12&0&0&&&\\
        0&0&0&0&0&0&0&0&0&&&\\ \hline
        0&0&0&0&0&0&0&0&0&&&\\
        \frac14&\frac14&0&0&0&-\frac12&0&0&0&&&\\
        1&1&0&0&0&1&0&0&0&&&\\
        \frac14&0&0&\frac14&0&2&-1&\frac12&0&&&\\ \hline
        &\frac16&\frac16&\frac23&0&-1&\frac43&\frac23&0&1&-\frac43&\frac13&0\\ \hline
        &&&&&2&-\frac43&-\frac23&0&-2&\frac53&-\frac23&1
\end{array}}.
\]}

The first scheme we will denote as
\texttt{SRK1W1}, the second --- as \texttt{SRK2W2}. The method
\texttt{SRK1W1} has strong order $(p_{d},p_{s}) = (2.0, 1.5)$, the 
method \texttt{SRK2W1} has strong order
$(p_{d},p_{s}) = (3.0, 1.5)$. Another scheme with strong order
$p_{s}=1.0$ can be found in~\cite{Kloeden_Platen}. This is the method with 
following Butcher table:
{\scriptsize%
\[
\text{\texttt{KlPl}: } {\renewcommand{\arraystretch}{1.1}%
\begin{array}{c|cc|cc|cc}
        0&0&0&0&0&&\\
        0&0&0&0&0&&\\ \hline
        0&0&0&0&0&&\\
        0&0&0&0&0&&\\ \hline
        0&1&0&1&0&-1&1\\
        & & &1&0&0&0\\
\end{array}}
\]}

On the figure.~\ref{fig:fig1} the local error of approximation of  SDE exact solution  
for logarithmic walk model is presented. The realisation of strong stochastic Runge--Kutta method 
with order $p_s = 1.5$ of strong convergence was used. 
It is interesting to note, that the approximation of
deterministic parts of the methods \texttt{SRK1W1} 
and \texttt{SRK2W1} don't essentially affect the error value.

\begin{minipage}[b]{0.45\linewidth}
\begin{figure}[H]
\centering
\includegraphics[width=1.0\linewidth]{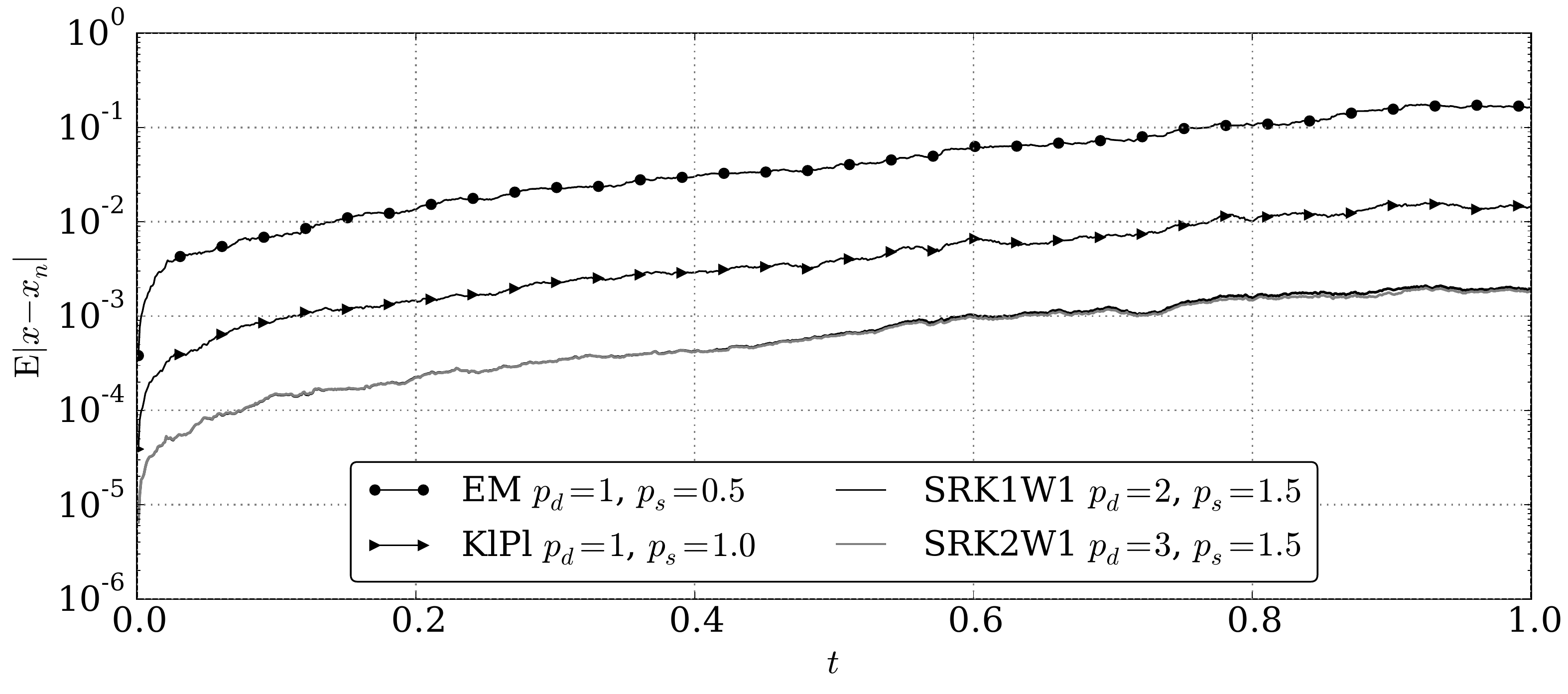}
\caption{The errors of methods for scalar process $W$ with
$h=10^{-3}$}
\label{fig:fig1}
\end{figure}
\end{minipage}
\hfill
\begin{minipage}[b]{0.45\linewidth}
\begin{figure}[H]
\centering
\includegraphics[width=1.0\linewidth]{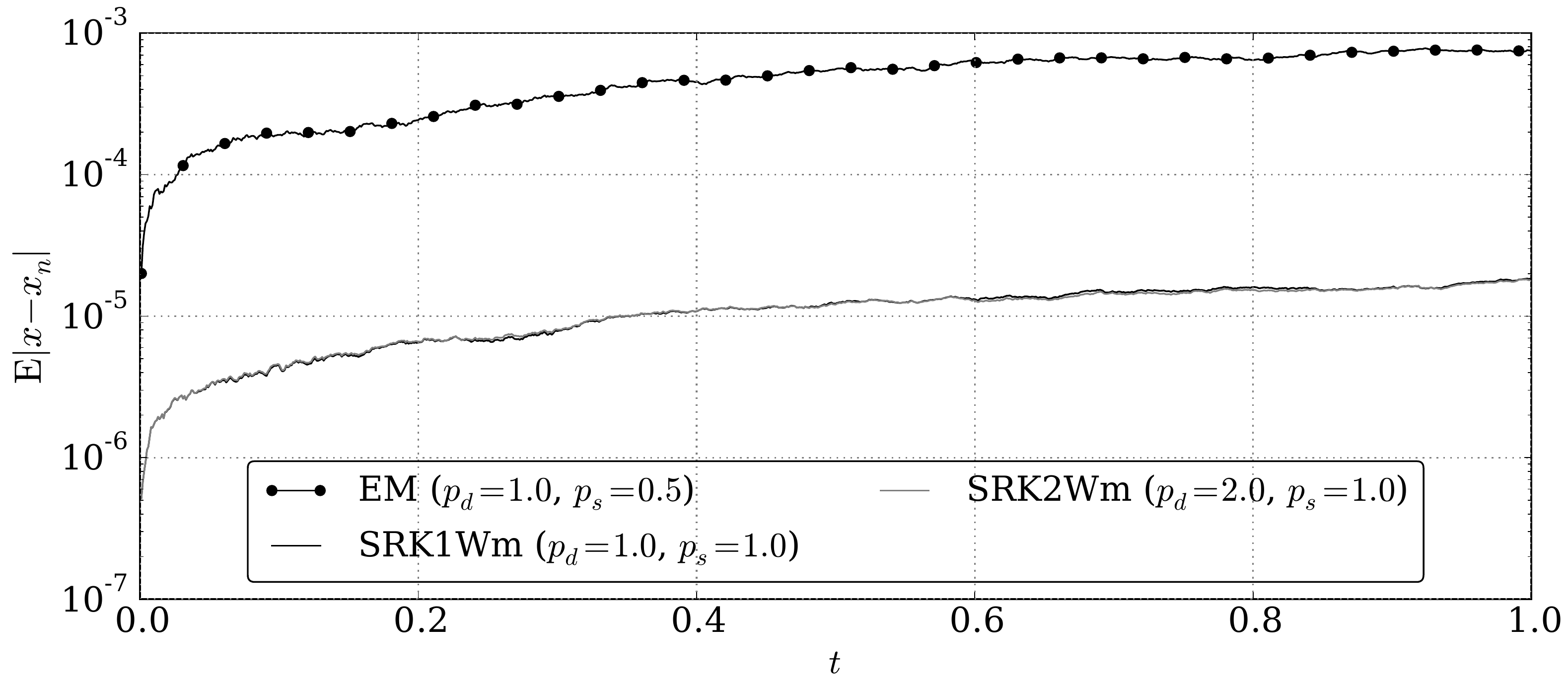}
\caption{The errors of methods for double process $W^{\alpha}$ 
with $h=10^{-3}$}
\label{fig:fig2}
\end{figure}
\end{minipage}

\subsection{Strong stochastic Runge--Kutta methods 
for SDE system with multidimensional Wiener process}

For It\^{o} SDE system with multidimensional Wiener process the 
stochastic Runge-Kutta scheme with strong order
$p_{s}=1.0$ can be obtained using single and double It\^{o} integrals~\cite{Rossler_2010}.

\begin{align*}
&X^{0i\alpha} = x^{\alpha}_{n} + \sum\limits^{s}_{j=1}A^{i}_{0j}f^{\alpha}(t_{n} + c^{j}_{0}h_{n}, X^{0j\beta})h_{n} + \sum\limits_{l=1}^{m}\sum\limits^{s}_{j=1}B^{i}_{0j}G^{\alpha}_{l}(t_{n} + c^{j}_{1}h_{n},X^{lj\beta})I^{l}(h_{n}),\\
&X^{ki\alpha} = x^{\alpha}_{n} + \sum\limits^{s}_{j=1}A^{i}_{1j}f^{\alpha}(t_{n} + c^{j}_{0}h_{n},X^{0j\beta})h_{n} + \sum\limits_{l=1}^{m}\sum\limits^{s}_{j=1}B^{i}_{1j}G^{\alpha}_{l}(t_{n} + c^{j}_{1}h_{n},X^{lj\beta})\dfrac{I^{lk}(h_{n})}{\sqrt{h_{n}}},\\
&x^{\alpha}_{n+1} = x^{\alpha}_{n} + \sum\limits^{s}_{i=1}a_{i}f^{\alpha}(t_{n} + c^{i}_{0}h_{n},X^{0i\beta})h_{n} + \sum\limits^{m}_{k=1}\sum\limits^{s}_{i=1}(b^1_{i}I^{k}(h_{n}) + b^2_{i}\sqrt{h_{n}})G^{\alpha}_{k}(t_{n} + c^{i}_{1}h_{n},X^{ki\beta}),
\end{align*}
where $n=0,1,\ldots,N-1$; $i=1,\ldots,s$; $\beta,k=1,\ldots,m$;
$\alpha = 1,\ldots,d$.

Butcher table~\cite{Rossler_2010} has the form:
\[ {\renewcommand{\arraystretch}{1.4}%
\begin{array}{c|c|c|c}
c^{i}_0 & A^{i}_{0j} & B^{i}_{0j} &\\ \hline
c^{i}_1 & A^{i}_{1j} & B^{i}_{1j} &\\ \hline
                                & a_{i} & b^{1}_{i} & b^{2}_{i} \\
\end{array}}.
\]

There are two realisations of this scheme for $s=3$ 
in Rößler preprint~\cite{Rossler_2010}:
{\scriptsize%
\[
\text{\texttt{SRK1Wm}: } {\renewcommand{\arraystretch}{1.1}%
\begin{array}{c|ccc|ccc|ccc}
        0&0&0&0&0&0&0&&&\\
        0&0&0&0&0&0&0&&&\\
        0&0&0&0&0&0&0&&&\\ \hline
        0&0&0&0&0&0&0&&&\\
        0&0&0&0&1&0&0&&&\\
        0&0&0&0&-1&0&0&&&\\ \hline
        &1&0&0&1&0&0&0&\frac{1}{2}&\frac{1}{2}
\end{array}}\;\;
\text{\texttt{SRK2Wm}: }
{\renewcommand{\arraystretch}{1.1}%
\begin{array}{c|ccc|ccc|ccc}
        0&0&0&0&0&0&0&&&\\
        1&1&0&0&0&0&0&&&\\
        0&0&0&0&0&0&0&&&\\ \hline
        0&0&0&0&0&0&0&&&\\
        1&1&0&0&1&0&0&&&\\
        1&1&0&0&-1&0&0&&&\\ \hline
        &\frac{1}{2}&\frac{1}{2}&0&1&0&0&0&\frac{1}{2}&-\frac{1}{2}
\end{array}}
\]}

The \texttt{SRK1Wm} method has the strong order
$(p_{d},p_{s}) = (1.0, 1.0)$ of convergence and the method \texttt{SRK2Wm} has the strong order
$(p_{d},p_{s}) = (2.0, 1.0)$ of convergence.

On figure~\ref{fig:fig2} the graph of strong convergence local error   
for Black--Scholes model is presented. It is also important to mention, that the 
deterministic part of methods \texttt{SRKp1W1} 
and \texttt{SRKp2W1} does not essentially  
influence the error value.

\subsection{Stochastic Runge--Kutta methodes with weak convergence  of $p=2.0$ order}

Numerical methods with weak convergence are the best for approximation of
characteristics of the distribution of the random process
$x^{\alpha}(t)$. Weak numerical method does not require information about
the exact trajectory of a Wiener process $W^{\alpha}_{n}$, and the random 
 variables for these methods can be generated on another probability
space. So we can use distribution which is easily generated on the computer.

For It\^{o} SDE system with multidimensional Wiener process 
the following stochastic Runge-Kutta scheme with weak order
$p_{s}=2.0$ is valid~\cite{Rossler_2010}.

\begin{align*}
& X^{0i\alpha} = x^{\alpha}_{n} + \sum\limits^{s}_{j=1}A_{0j}^{i}f^{\alpha}(t_{n} + c^{j}_{0}h_{n}, X^{0j\beta})h_{n} + \sum\limits^{s}_{j=1}\sum\limits^{m}_{l=1}B_{0j}^{i}G^{\alpha}_{l}(t_{n}+c^{j}_{1}h_{n}, X^{lj\beta})\hat{I}^{l},\\
& X^{ki\alpha} = x^{\alpha}_{n} + \sum\limits^{s}_{j=1}A^{i}_{1j}f^{\alpha}(t_{n}+c^{j}_{0}h_{n}, X^{0j\beta})h_{n} + \sum\limits^{s}_{j=1}B^{i}_{1j}G^{\alpha}_{k}(t_{n} + c_{1}^{j}h_{n}, X^{kj\beta})\sqrt{h_{n}},\\
& \widehat{X}^{ki\alpha} = x^{\alpha}_{n} + \sum\limits^{s}_{j=1}A^{i}_{2j}f^{\alpha}(t_{n}+c^{j}_{0}h_{n}, X^{0j\beta})h_{n} + \sum\limits^{s}_{j=1}\sum\limits^{m}_{l=1,l\neq k}B^{i}_{2j}G^{\alpha}_{l}(t_{n} + c_{1}^{j}h_{n}, X^{lj\beta})\frac{\hat{I}^{kl}}{\sqrt{h_{n}}},\\
& x^{\alpha}_{n+1} = x^{\alpha}_{n} + \sum\limits^{s}_{i=1}a_{i}f^{\alpha}(t_{n}+c^{i}_{1}, X^{ki\beta})h_{n} + \sum\limits^{s}_{i=1}\sum\limits^{m}_{k=1}\left(b^{1}_{i}\hat{I}^{k} + b^{2}_{i}\frac{\hat{I}^{kk}}{\sqrt{h_{n}}}\right) G^{\alpha}_{k}(t_{n}+c^{i}_{1}h_{n}, X^{ki\beta}) + \\
& +  \sum\limits^{s}_{i=1}\sum\limits^{m}_{k=1}\left(b^{3}_{i}\hat{I}^{k} + b^{4}_{i}\sqrt{h_{n}}\right)G^{\alpha}_{k}(t_{n}+c^{i}_{2}h_{n}, \widehat{X}^{ki\beta}).
\end{align*}

Butcher table~\cite{Rossler_2010} in this case appears as follows:
\[
{\renewcommand{\arraystretch}{1.5}%
\begin{array}{c|c|c|c}
c^{i}_{0} & A^{i}_{0j} & B^{i}_{0j} &\\ \hline
c^{i}_{1} & A^{i}_{1j} & B^{i}_{1j} &\\ \hline
c^{i}_{2} & A^{i}_{2j} & B^{i}_{2j} &\\ \hline
                                        & a_{i} & b^{1}_{i} & b^{2}_{i}\\ \hline
                                        &  & b^{3}_{i} & b^{4}_{i}\\
\end{array}}
\] {\scriptsize%
\[
{\renewcommand{\arraystretch}{1.5}%
\begin{array}{c|ccc|ccc|ccc}
        0&0&0&0&0&0&0&&&\\
        1&1&0&0&\frac13&0&0&&&\\
        \frac{5}{12}&\frac{25}{144}&\frac{35}{144}&0&-\frac56&0&0&&&\\ \hline
        0&0&0&0&0&0&0&&&\\
        \frac14&\frac14&0&0&\frac12&0&0&&&\\
        \frac14&\frac14&0&0&-\frac12&0&0&&&\\ \hline
        0&0&0&0&0&0&0&&&\\
        0&0&0&0&1&0&0&&&\\
        0&0&0&0&-1&0&0&&&\\ \hline
        &\frac{1}{10}&\frac{3}{14}&\frac{24}{35}&1&-1&-1&0&1&-1\\ \hline
        &&&&\frac12&-\frac14&-\frac14&0&\frac12&-\frac12\\
\end{array}}\quad
{\renewcommand{\arraystretch}{1.5}%
\begin{array}{c|ccc|ccc|ccc}
        0&0&0&0&0&0&0&&&\\
        1&1&0&0&1&0&0&&&\\
        0&0&0&0&0&0&0&&&\\ \hline
        0&0&0&0&0&0&0&&&\\
        1&1&0&0&1&0&0&&&\\
        1&1&0&0&-1&0&0&&&\\ \hline
        0&0&0&0&0&0&0&&&\\
        0&0&0&0&1&0&0&&&\\
        0&0&0&0&-1&0&0&&&\\ \hline
        &\frac{1}{2}&\frac{1}{2}&0&\frac12&\frac14&\frac14&0&\frac12&-\frac12\\ \hline
        &&&&-\frac12&\frac14&\frac14&0&\frac12&-\frac12\\
\end{array}}.
\]}

In the weak stochastic Runge--Kutta scheme the following
random variables are used:
\[
\hat{I}^{kl} = \left\{
\begin{aligned}
&\dfrac{1}{2}(\hat{I}^{k}\hat{I}^{l} - \sqrt{h}_{n}\tilde{I}^{k}),\; k<l,\\
&\dfrac{1}{2}(\hat{I}^{k}\hat{I}^{l} + \sqrt{h}_{n}\tilde{I}^{l}),\; l<k,\\
&\dfrac{1}{2}((\hat{I}^{k})^2 - h_{n}).\; k=l.
\end{aligned}
\right.
\]
The random variable $\hat{I}^{k}$ has three-points distribution, it can take on
three fixed values: $ \{-\sqrt{3h_{n}}, 0, \sqrt{3h_{n}}\}$ with probabilities
$1/6$, $2/3$ and $1/6$ respectively. The variable  $\tilde{I}^{k}$
has two-points distribution with two values $\{-\sqrt{h_{n}}, \sqrt{h_{n}}\}$ 
and probabilities $1/2$ and $1/2$.

\subsection{Strong and weak convergence error}

To calculate the strong convergence error we have to generate a single 
Wiener process trajectory, and then we need to use this trajectory as driving Wiener 
process in order to calculate the exact solution, as well as approximate solution. After
that we can find local $\|\mathbf{x}(t_{n}) - \mathbf{x}_{n}\|$, $n=1,\ldots,N$ 
or global $\|\mathbf{x}(T) - \mathbf{x}_{N}\|$ approximation errors.

Weak error calculation is not such a simple task, because we have to use Monte-Carlo
        method for it. For the brief reference see~\cite{Andreas_2003}.

\section{Sage \texttt{sde} module reference}

\begin{figure}
                \centering
                \includegraphics[width=0.7\linewidth]{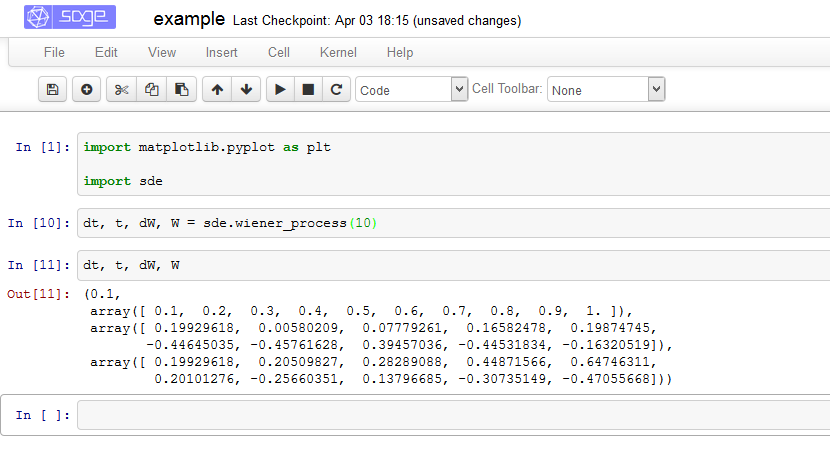}
                \caption{The usage of \texttt{sde} module in Sage \texttt{ipython notebook} mode}
                \label{fig:sage}
        \end{figure}

Our library is a common python module. To connect it to Sage users simply 
perform a standard command \spverb|import sde|. Fig.~\ref{fig:sage} shows 
the Sage notebook in ipython mode (\texttt{--notebook=ipython} 
command line option)

The library contains a number of functions for internal
use. The names of these functions begin with the double bottom
underscore, as required by the style rules for python code PEP8.
\begin{itemize}
\item \spverb|(dt, t) = __time(N, interval=(0.0, 1.0))| --- function divides 
the time interval \texttt{interval} into
  \texttt{N}subintervals and returns \texttt{numpy} array \texttt{t} with 
        step \texttt{dt};

\item \spverb|(dW, W) = __scalar_wiener_process(N, dt, seed=None)| ---
  function generates a trajectory of a scalar Wiener process
  \texttt{W} from \texttt{N} subintervals with step \texttt{dt};

\item
  \spverb|(dW, W) = __multidimensional_wiener_process(N, dim, dt, seed=None)|
  --- similar function for generating multidimensional (\texttt{dim} dimensions) 
        Wiener process;

\item
  \spverb|(dW, W) = __cov_multidimensional_wiener_process(N, dim, dt, seed=None)|
  --- another function for generating multidimensional (\texttt{dim} dimensions) 
        Wiener process, which uses \texttt{numpy.multivariate\_normal}. 
        $W^1,W^2,\ldots,W^m$ processes;

\item
  \spverb|(dt, t, dW, W) = wiener_process(N, dim=1, interval=(0.0, 1.0), seed=None)|
  --- the main function which should be used to generate
   Wiener process. Positional argument $N$ denotes the number of points
        on the interval \texttt{interval} (default $[0,1]$). 
        Argument \texttt{dim} defines a dimension of Wiener process. 
        Function returns sequence of four elements \texttt{dt, t, dW, W}, where
  \texttt{dt} is a step, ($\Delta t_{i} = h = \mathrm{const}$),
  \texttt{t} is one-dimensional \texttt{numpy}-array of time points 
        $t_{1},t_{2},\ldots,t_{N}$; \texttt{dW}, \texttt{W}
   are also $N\times m$ dimensional \texttt{numpy}-arrays,
   consisting of increments $\Delta W^{\alpha}_{i}$ and trajectory points
         $W^{\alpha}_{i}$, where $i=1,\ldots,N;\;\alpha=1,\ldots,m$.
        
\item
  \spverb|__strong_method_selector(name) __weak_method_selector(name)|
  --- set Butcher table for specific method.
\end{itemize}

        The following set of functions are made for It\^{o} integrals 
        approximation in strong Runge-Kutta schemas. All of them get an array 
        of increments \texttt{dW} as positional argument. Step size \texttt{h} is 
        the second argument. All functions return a list of integral approximations
        for each point of driving process trajectory. For multiple integrals this 
        list contains arrays.
\begin{itemize}
\item \spverb|Ito1W1(dW)| --- function generates values of single It\^{o} integral 
for scalar driving Wiener process;
\item \spverb|Ito2W1(dW, h)| --- function generates values of double It\^{o} integral 
for scalar driving Wiener process;
\item \spverb|Ito1Wm(dW, h)| --- function generates values of single It\^{o} integral 
for multidimensional driving Wiener process;
\item \spverb|Ito2Wm(dW, h, n)| --- function approximate values of double It\^{o} integral. 
Function returns a list of arrays $I^{ij}$ for each trajectory step. 
Argument \texttt{n} denotes the number of terms in infinite series $\mathbf{A}(h)$;
\item \spverb|Ito3W1(dW, h)| --- function generates values of triple It\^{o} integral 
for scalar driving Wiener process.
\end{itemize}

 Now we will describe core functions that implement numerical methods with
strong convergence.
\begin{itemize}
\item
  \spverb|EulerMaruyama(f, g, h, x_0, dW) EulerMaruyamaWm(f, g, h, x_0, dW)|
  --- EulerMaruyama method for scalar and multidimensional Wiener
   processes, where \texttt{f} --- a drift vector, \texttt{g} --- a diffusion matrix;
        
\item \spverb|strongSRKW1(f, g, h, x_0, dW, name='SRK2W1')| ---
  function for SDE integration with scalar Wiener process, where
  \texttt{f(x)} and \texttt{g(x)} are the same as in above functions;
  \texttt{h} --- a step, \texttt{x\_0} --- an initial value,
  \texttt{dW} --- an array of Wiener process increments \texttt{N};
  argument \texttt{name} can take values \texttt{'SRK1W1'} and
  \texttt{'SRK2W2'};

\item \spverb|oldstrongSRKp1Wm(f, G, h, x_0, dW, name='SRK1Wm')| ---
  stochastic Runge--Kutta method with strong convergence order $p = 1.0$ for
  multidimensional Wiener process. This function is left only for 
        performance testing for nested loops realisation of 
        multidimensional arrays convolution ;

\item \spverb|strongSRKp1Wm(f, G, h, x_0, dW, name='SRK1Wm')| ---
  stochastic Runge--Kutta method with strong convergence order $p = 1.0$ for
  multidimensional Wiener process. We use NumPy method
  \texttt{numpy.einsum} because it gives sufficient performance goal.
\end{itemize}

The next group of functions implements methods with weak convergence.
\begin{itemize}
\item \spverb|n_point_distribution(values, probabilities, shape)| ---
  n-points distribution, where \spverb|values| --- an array of random variable values, 
        \spverb|probabilities| --- an array of probabilities, \spverb|shape| --- 
        arrays shape;
        
\item \spverb|weakIto(h, dim, N)| --- It\^{o} integrals replacement for 
weak convergence;

\item \spverb|weakSRKp2Wm(f, G, h, x_0, dW, name='SRK1Wm')| ---
  stochastic Runge--Kutta method  for multidimensional
        Wiener process with order $p = 2.0$.
\end{itemize}

As mentioned above, in order to apply stochastic numerical methods with weak convergence we
have to use Monte-Carlo method. Thus, we must lunch a series of multiple integration
of SDE using weak Runge--Kutta method. This is very expensive computational process,
 so the parallel computing is needed. We use standard python \spverb|multiprocessing| 
 module for this purpose. The module has a set of functions for parallel process 
 creation and task sharing. Since every computation is done completely independent, 
 the processes do not require to share any data with each other.
 
 It is important to mention that \spverb|multiprocessing| use \spverb|fork| system call
 to create processes. That's why all processes inherit the same number generator seed
  so they create exactly similar Wiener processes trajectories. To overcome 
        this we define optional \spverb|seed| parameter for all functions, which use random number 
        generator. It takes integer values and is used for random number generator initialization.
         During parallel computing a computation sequence number can be utilized as seed|parameter.
        More details and couple examples are available on repository \url{https://bitbucket.org/mngev/sde-numerical-integrators}.

\section{Stochastic ``predator-pray'' model}

We use our library to calculate the numerical approximation
for stochastic ``predators-pray''
model~\cite{kulyabov:2013:conf:mmcp,kulyabov:2014:icumt-2014:p2p,ef-kor-gev-kul-sev:vestnik-miph:2014-3}.

``Predator-pray'' model is defined by SDE with the following drift vector 
$f$ and diffusion matrix $G$

\begin{equation}
f(x,y) = \left(
  \begin{array}{c}
    k_{1}ax - k_{2}xy\\
    k_{2}xy - k_{3}y
  \end{array}
\right) \qquad G(x,y) = \left(
  \begin{array}{cc}
    k_{1}ax+k_{2}xy & -k_{2}xy\\
    -k_{2}xy & k_{2}xy + k_{3}y
  \end{array}
\right)^{\frac{1}{2}},
\end{equation}

where $x$, $y$ are prays and predators populations respectively, 
$a$ denotes prays growth, $k_{1}$ denotes prays competition, $k_{2}$ is 
a rate of predation, $k_{3}$ is predator's death rate. We choose parameters values 
$a, k_1, k_2, k_3 = (0.5, 3.0, 0.05, 2.5)$ and initial point $(x_0, y_0) = (50.0, 30.0)$ to be 
closely related to the stationary point.

For stochastic ``predator-pray'' model the invariant $\mathrm{d}I(x,y)$:
\[
\mathrm{d}I(x,y) = \frac12 \left( \frac{ak_{1}k_{3}}{x} +
  \frac{k_{2}k_{3}y}{x} + \frac{ak_{1}k_{2}x}{y} +
  \frac{ak_{1}k_{3}}{y} \right),
\]
plays an important role. This invariant characterizes the phase volume 
of the system and it grows during system evolution~\cite{kulyabov:2013:conf:mmcp}.
We use strong and weak methods to check model properties.

We use weak Runge--Kutha method to calculate 500 exact realisations
of SDE numerical solutions. Based on this data we find different statistical
characteristics, such as median and percentiles. Time evolution of these 
characteristics illustrates key fiches of stochastic ``predator-pray'' model.
Thus from fig.~\ref{fig:prepre4} we can see that the phase volume increases over time, 
also fig.~\ref{fig:prepre5} illustrates the same behaviour of $\mathrm{d}I$ value.

\begin{figure}
  \centering
  \includegraphics[width=0.6\linewidth]{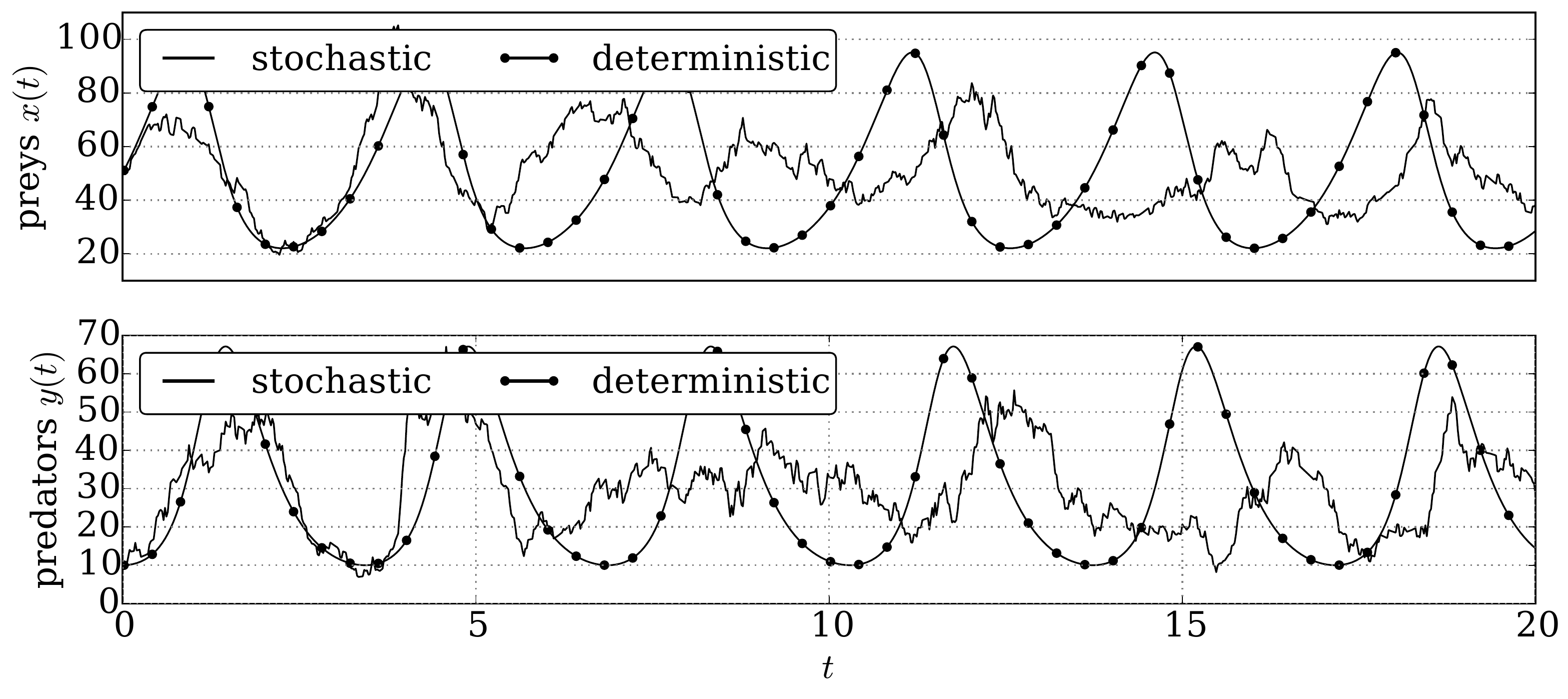}
  \caption{``Predator-pray'' model numerical solutions (strong Runge-Kutta method)}
  \label{fig:prepre1}
\end{figure}

\begin{minipage}[b]{0.45\linewidth}
\begin{figure}[H]
  \centering
  \includegraphics[width=1.0\linewidth]{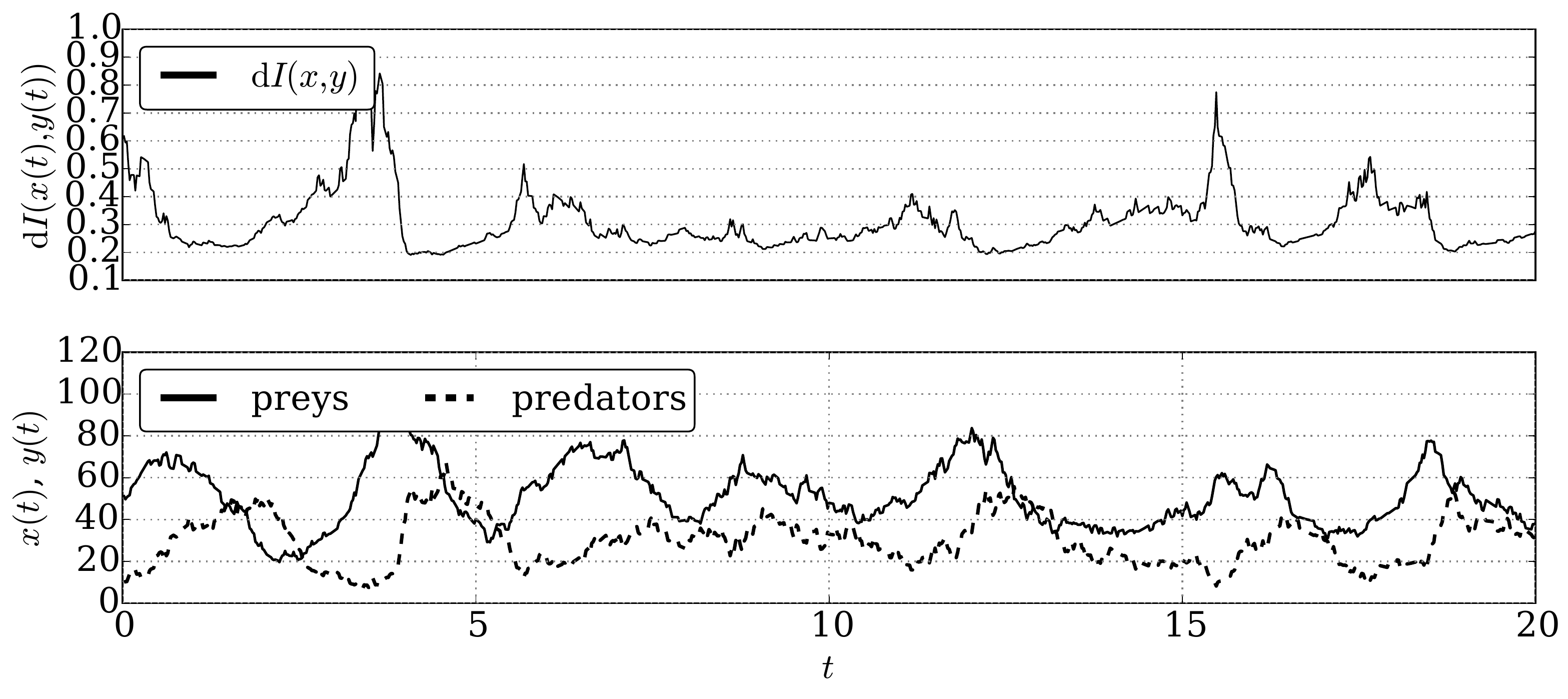}
  \caption{Invariant $\mathrm{d}I$ time evolution
        for process trajectory}
  \label{fig:prepre2}
\end{figure}
\end{minipage}
\hfill
\begin{minipage}[b]{0.45\linewidth}
\begin{figure}[H]
  \centering
  \includegraphics[width=1.0\linewidth]{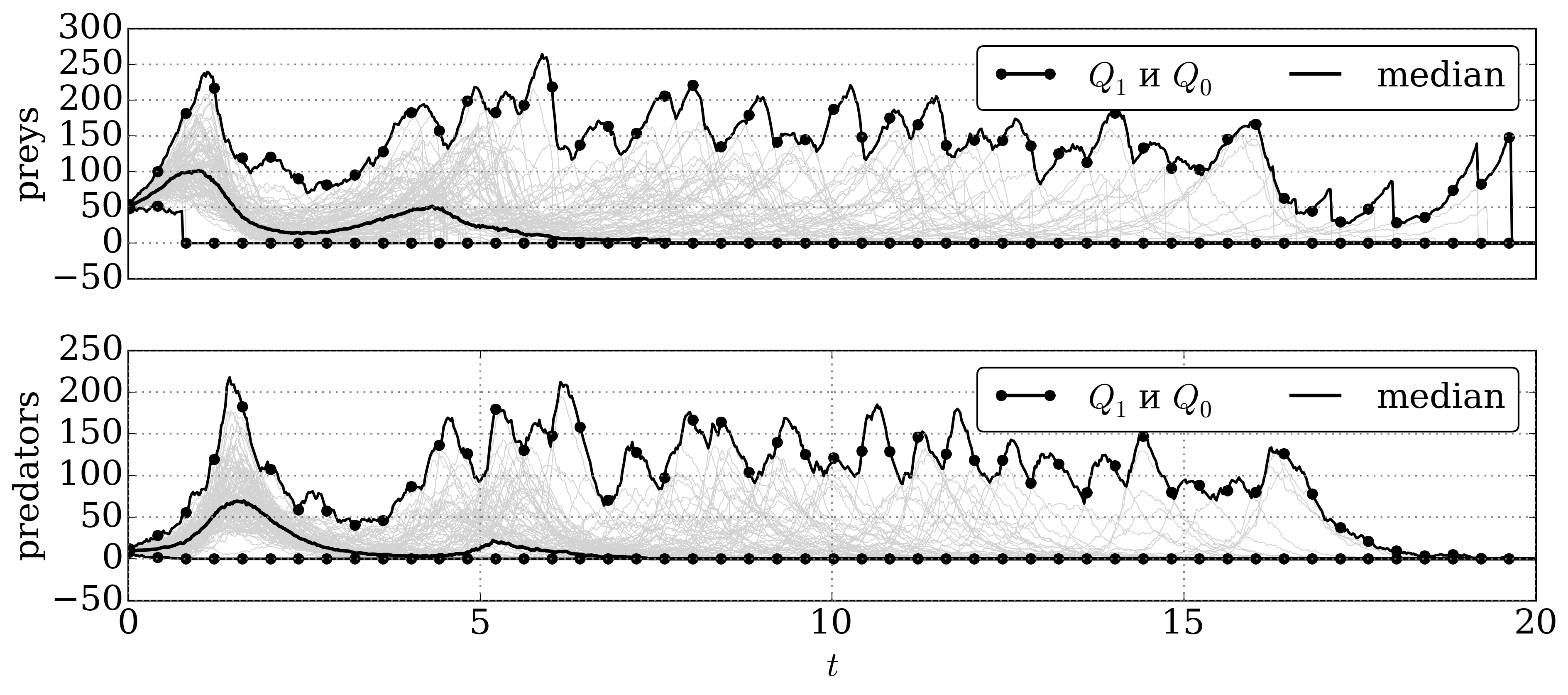}
  \caption{Multiple trajectories (gray), 
        $Q_{1}, Q_{0}$ --- percentiles and median.}
  \label{fig:prepre3}
\end{figure}
\end{minipage}

\begin{minipage}[b]{0.45\linewidth}
\begin{figure}[H]
  \centering
  \includegraphics[width=1.0\linewidth]{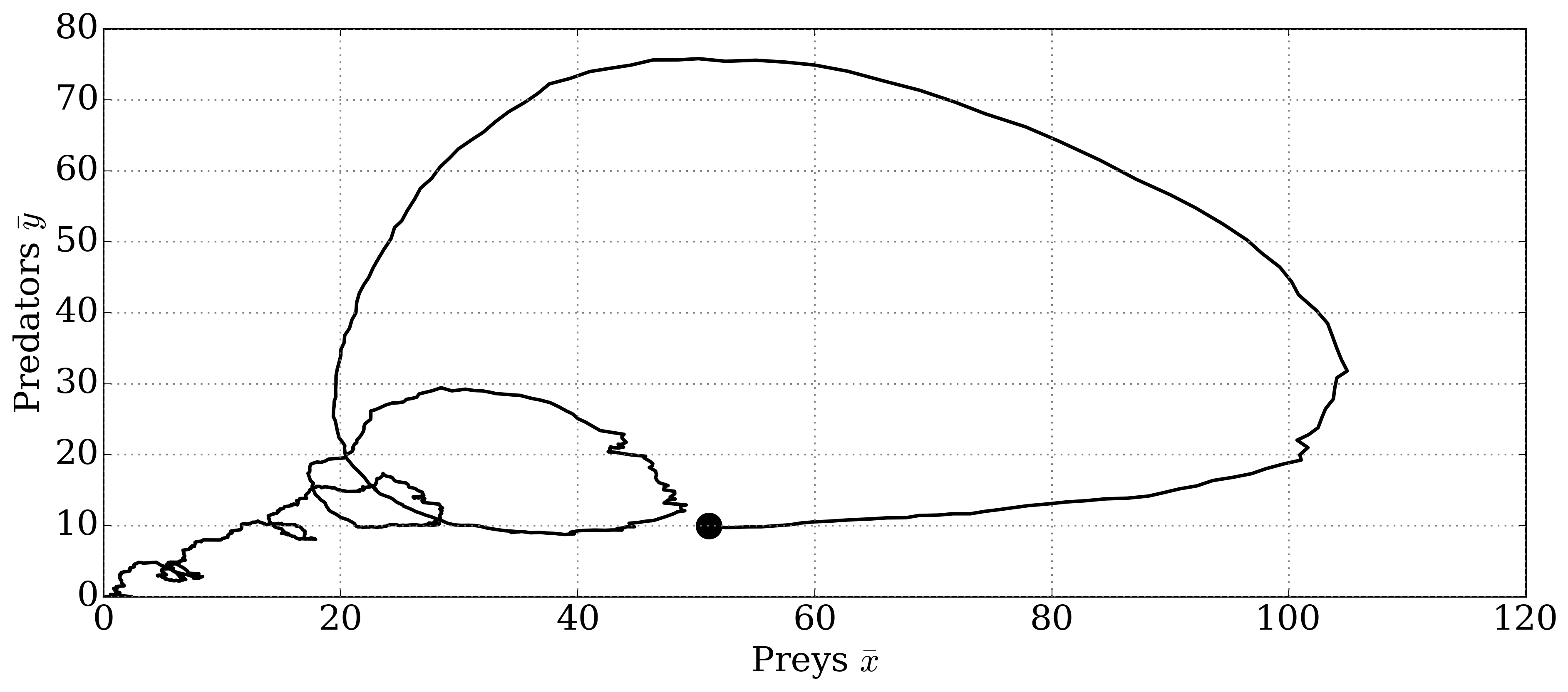}
  \caption{The phase portrait of the sample means of predators and prey quantities,
        based on 500 trajectories obtained by Runge--Kutta method with weak convergence.}
  \label{fig:prepre4}
\end{figure}
\end{minipage}
\hfill
\begin{minipage}[b]{0.45\linewidth}
\begin{figure}[H]
  \centering
  \includegraphics[width=1.0\linewidth]{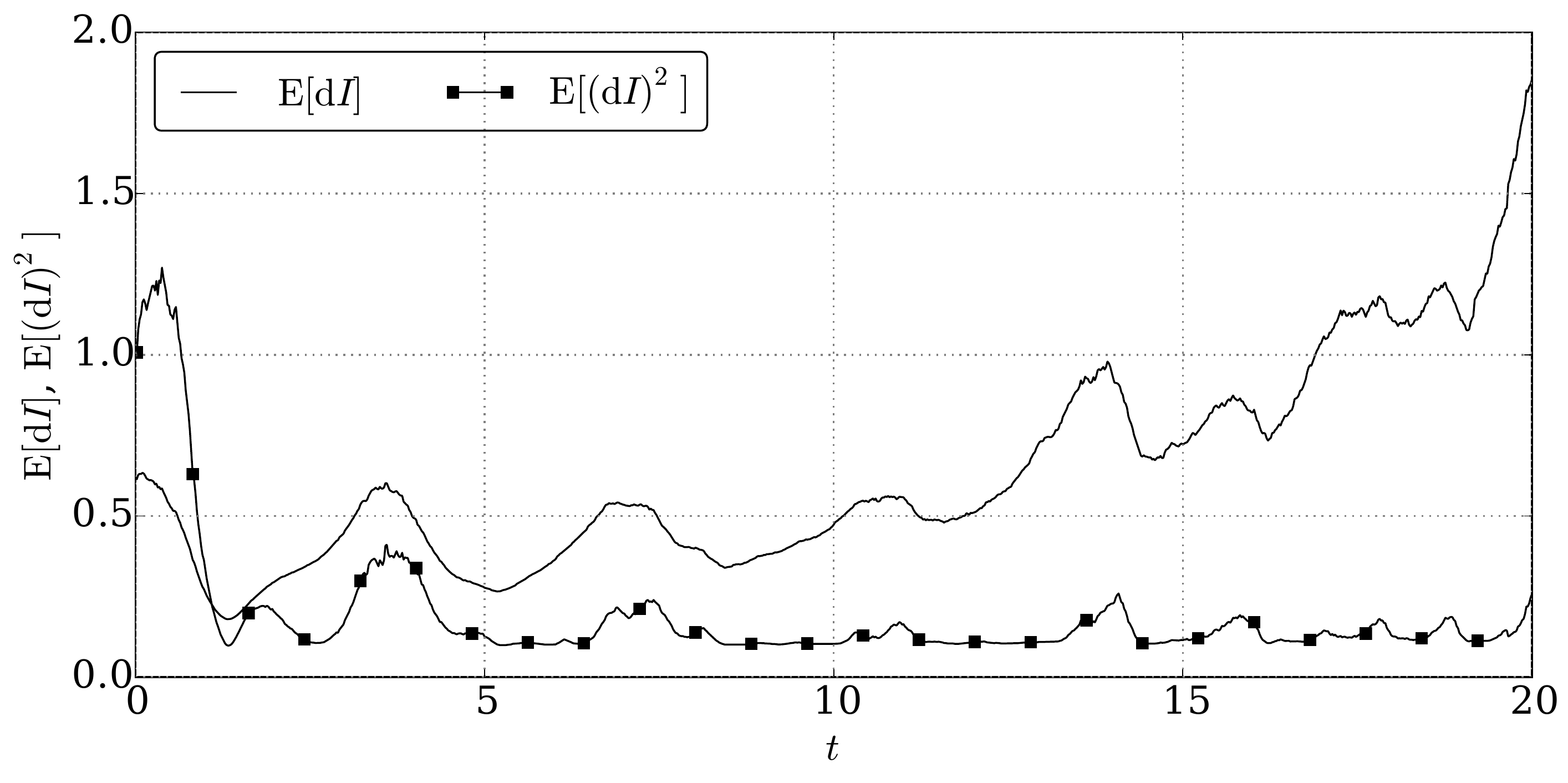}
  \caption{First and second empirical moments
    $\mathrm{d}I$ based on 500 trajectories obtained 
                by Runge--Kutta method with weak convergence.}
  \label{fig:prepre5}
\end{figure}
\end{minipage}

Another key difference between deterministic and stochastic models is the tendency 
of imminent death of population during evolution. On the fig.~\ref{fig:prepre3}
grey colour denotes all trajectories realisations. It is seen that trajectories 
intersect the x-axis, thus the population of prays/predators dies. It makes stochastic
model more realistic, because  in the deterministic case the population death is not possible.

\section{Conclusions}
\label{sec:conclusion}

Some realizations of stochastic Runge-Kutta methods were considered in this article.
The authors gradually developed and refined the library by adding a new 
functionalities and optimizing existing ones. To date, 
the library uses numerical methods by the Rößler's article~\cite{Rossler_2010}, 
as the most effective of the currently known to the authors. 
However, the basic functions \texttt{strongSRKW1} and \texttt{weakSRKp2Wm}  
are written in accordance with the general algorithm and can use any Butcher table with appropriate 
staging. This allows any user of the library to extend functionality by adding new methods.

Additional examples of library usage and of all files source codes  are available at 
 \url{https://bitbucket.org/mngev/sde-numerical-integrators}.
 The authors suppose to maintain the library by adding new functions.

\begin{acknowledgments}

The work is partially supported by RFBR grants No's 14-01-00628,
15-07-08795, and 16-07-00556.

Calculations were carried out on computer cluster ``Felix'' (RUDN) and
Heterogeneous computer cluster ``HybriLIT'' (Multifunctional
center storage, processing and analysis of data JINR).

Published in:
M.~N. Gevorkyan, T.~R. Velieva, A.~V. Korolkova, D.~S. Kulyabov, L.~A.
Sevastyanov, Stochastic Runge–Kutta Software Package for Stochastic
Differential Equations, in: Dependability Engineering and Complex Systems,
Vol. 470, Springer International Publishing, 2016, pp. 169--179.
\href{http://dx.doi.org/10.1007/978-3-319-39639-2_15}{doi:10.1007/978-3-319-39639-2\_15}.

Sources:
\url{https://bitbucket.org/yamadharma/articles-2015-rk-stochastic}

\end{acknowledgments}

\bibliographystyle{elsarticle-num}

\bibliography{bib/rk-stoch-realis/bib,bib/rk-stoch-realis/self}

\end{document}


\graphicspath{{image/rk-stoch-realis/en/}}

\title{Программный комплекс для решения стохастических
  дифференциальных уравнений стохастическим методом Рунге--Кутты}

\author{М. Н. Геворкян}
\email{mngevorkyan@sci.pfu.edu.ru} 
\affiliation{Кафедра прикладной информатики и теории вероятностей,\\
  Российский университет дружбы народов,\\
  ул. Миклухо-Маклая, д.6, Москва, Россия, 117198}

\author{Т. Р. Велиева}
\email{trvelieva@gmail.com}
\affiliation{Кафедра прикладной информатики и теории вероятностей,\\
  Российский университет дружбы народов,\\
  ул. Миклухо-Маклая, д.6, Москва, Россия, 117198}

\author{А. В. Королькова}
\email{avkorolkova@gmail.com} 
\affiliation{Кафедра прикладной информатики и теории вероятностей,\\
  Российский университет дружбы народов,\\
  ул. Миклухо-Маклая, д.6, Москва, Россия, 117198}

\author{Д. С. Кулябов}
\email{yamadharma@gmail.com}
\affiliation{Кафедра прикладной информатики и теории вероятностей,\\
  Российский университет дружбы народов,\\
  ул. Миклухо-Маклая, д.6, Москва, Россия, 117198}
\affiliation{Лаборатория информационных технологий,\\
Объединённый институт ядерных исследований,\\
ул. Жолио-Кюри 6, Дубна, Московская область, Россия, 141980}

\author{Л. А. Севастьянов}
\email{leonid.sevast@gmail.com}
\affiliation{Кафедра прикладной информатики и теории вероятностей,\\
  Российский университет дружбы народов,\\
  ул. Миклухо-Маклая, д.6, Москва, Россия, 117198}
\affiliation{Лаборатория теоретической физики,\\
  Объединённый институт ядерных исследований,\\
  ул. Жолио-Кюри 6, Дубна, Московская область, Россия, 141980}

\begin{abstract}

  В результате применения методики построения стохастических моделей
  одношаговых процессов был получен спектр моделей, реализованный в
  виде самосогласованных дифференциальных уравнений. Это уравнения в
  частных производных (основное кинетическое уравнение, уравнение
  Фоккера--Планка) и стохастические дифференциальные уравнения
  (уравнение Ланжевена). Однако аналитические методы не всегда
  позволяют провести их адекватное исследование.  Предлагается при
  исследовании данных уравнений использовать комбинированный
  аналитико-числовой подход. Для этого численная часть реализуется в
  рамках системы символьных вычислений.  При численном исследовании
  стохастических дифференциальных уравнений в форме Ланжевена
  предлагается применять стохастические методы Рунге--Кутты. В рамках
  данного подхода создан программный комплекс на базе метасистемы
  аналитических вычислений Sage. Для верификации использованы модели
  логарифмического блуждания и двумерная модель Блека--Шоулза. Для
  иллюстрации использована стохастическая модель типа
  <<хищник--жертва>>.  Продемонстрировано удобство применения
  комбинированного численно-аналитического подхода.








\end{abstract}

  \keywords{метод Рунге--Кутты; стохастические дифференциальные
    уравнения; модель <<хищник--жертва>>; символьные методы в биологии; системы компьютерной
    алгебры; система Sage}

\maketitle

\section{Введение}

Учёт стохастических свойств при моделировании динамических систем
позволяет адекватнее описывать реальные процессы. Стохастические
модели применяются в химической кинетике, гидродинамике, популяционной
динамике, эпидемиологии, обработке и фильтрации сигналов, экономике и
финансовой математике, а также в различных разделах
физики~\cite{Kloeden_Platen}. Основным математическим аппаратом для
таких моделей являются стохастические дифференциальные уравнения
(СДУ).

Даже в случае детерминированных дифференциальных уравнений точное
аналитическое решение может быть получено лишь для ограниченного
класса уравнений. На практике этот недостаток компенсируется
многочисленными и хорошо разработанными численными
методами~\cite{Butcher_2003,Hairer:1::ru}. В случае СДУ
значение численных методов повышается, так как точные аналитические
решения получены для очень небольшого числа стохастических
моделей~\cite{Kloeden_Platen,Andreas_2003}.

По сравнению с численными методами для обыкновенных дифференциальных
уравнений (ОДУ), численные методы для СДУ развиты на порядок
слабее. Основных причин для этого две: сравнительная новизна данной
темы и значительно более сложный математический аппарат. Направления
исследований и поиск новых численных методов для стохастического
случая в целом совпадает с классическим
случаем~\cite{Burrage_1996,Burrage_1997,Burrage_1998,Burrage_2000,Rossler_2010,Andreas_2003}. Наглядное
представление об этих направлениях даёт схема, предложенная в книге
Бутчера~\cite{Butcher_2003}.


Точность аппроксимации можно повышать, используя в численной схеме
значения производных от входящих в уравнение коэффициентов диффузии и
сноса (\emph{мультидифференцируемость}), вычисление каждого шага в
несколько стадий (\emph{многостадийность}), а также 
(\emph{многошаговость}).

В случае детерминированных дифференциальных уравнений для программной
реализации наиболее удобны многостадийные явные методы (методы типа
Рунге-Кутты), так как они сводятся к последовательному вычислению ряда
величин с помощью явных формул. Естественно, что одним из основных
направлений исследования стало распространение теории методов
Рунге-Кутты на случай СДУ.

Целью данной статьи является описание авторского программного
комплекса для численного решения СДУ стохастическими методами
Рунге-Кутты с попутным описанием всех используемых
алгоритмов. Библиотека предназначена для CAS Sage~\cite{sage} и
написана на языке \texttt{python} с использованием модулей
\texttt{numpy} и \texttt{scipy}.

Для программирования с использованием стохастических численных методов
необходимо было решить ряд \textbf{задач} по реализации
вспомогательных алгоритмов, таких как генерирование траекторий
винеровского процесса и многоточечных распределений, аппроксимация
кратных стохастических интегралов, тестирование сильной и слабой
сходимости численных методов. Только после их решения стало возможным
реализовать непосредственно сами численные алгоритмы стохастического
метода Рунге--Кутты.

Авторы столкнулись с необходимостью программной реализации
стохастических численных методов в ходе работы над методикой
согласованного введения стохастики в детерминированные
модели~\cite{kulyabov:2013:conf:mmcp,kulyabov:2014:icumt-2014:p2p,ef-kor-gev-kul-sev:vestnik-miph:2014-3}. В
качестве примеров для применения методики были взяты модели типа
хищник-жертва. Следовало верифицировать полученные результаты
(стохастические модели в форме СДУ) при помощи численных
экспериментов. Сходная задача возникла при работе над стохастическими
моделями дисциплин обслуживания очередей типа RED на
маршрутизаторах~\cite{kulyabov:2014:icumt-2014:gns3}.

Необходимо было сравнить результаты численного решения 
математической модели с результатами имитационного моделирования.

\subsection{Структура работы}

В первой части статьи изложен минимум необходимых теоретических
сведений: определение скалярного и многомерного процессов Винера,
определение скалярного и многомерного СДУ в форме Ито, определение
сильной и слабой сходимости аппроксимирующей функции, матричные
формулы для аппроксимации двукратных интегралов Ито.

Вторая часть статьи посвящена описанию стохастических численных схем
Рунге--Кутты с сильным и слабым порядками сходимости, как для
скалярного, так и для многомерного винеровских процессов. 

В третей части дается справка по функциям библиотеки, которые реализуют 
вышеописанные численные схемы.

Завершает изложение пример использования реализованных алгоритмов для
нахождения сильной и слабой аппроксимации решения стохастической
модели хищник-жертва. На основе проведенных численных экспериментов
делаются некоторые выводы о ключевых отличиях стохастической модели от
детерминированной.

Статью можно рассматривать как практическое введение в область
стохастических численных методов. Приводимые сведения будут полезны
при реализации стохастических численных схем на других языках
программирования.

\subsection{Выбор языка программирования для реализации стохастических
  численных методов}

При выборе системы компьютерной алгебры для реализации стохастических
численных методов Рунге--Кутты авторы руководствовались следующими
требованиями:
\begin{itemize}
\item необходимы расширенные средства для работы с многомерными
  массивами (до четырех размерностей) с большим количеством элементов;
\item критична возможность параллельного выполнения отдельных функций
  и участков кода ввиду необходимости применения метода Монте-Карло с
  большим числом независимых испытаний;
\item желательна возможность генерирования массивов случайных чисел;
\item программное обеспечение должно быть открытым и бесплатным для
  некоммерческого использования.
\end{itemize}

В целом всем этим требованием удовлетворяет CAS Sage, написанная на
языке \texttt{python}. Все необходимые функции для работы с массивами
реализованы в библиотеке \texttt{numpy}, а в библиотеке \texttt{scipy}
присутствуют широкий набор средств для генерирования случайных
чисел. Кроме того библиотека \texttt{numpy} обеспечивает высокую
производительность.

\section{Основные понятия теории стохастических дифференциальных
  уравнений}
\subsection{Случайный процесс Винера}

Здесь излагаются лишь самые необходимые сведения. Для быстрого
введения в область СДУ см.~\cite{Higham_2001,Malham_2010}, а для
более обстоятельного знакомства
см.~\cite{Kloeden_Platen,Platen_Liberati_2010,Oksendal::ru}.

\begin{thedefinition}
  Случайный процесс $W(t)$, $t \geqslant 0$ называется скалярным
  \emph{процессом Винера}, если выполняются следующие условия:
  \begin{itemize}
  \item $\mathrm{P}\{W(0)=0\} = 1$, иначе говоря $W(0) = 0$ почти
    наверное;
  \item $W(t)$ --- процесс с независимыми приращениями, то есть
    $\{\Delta W_{i}\}^{N-1}_{0}$~--- независимые случайные величины;
    $\Delta W_{i} = W(t_{i+1}) - W(t_{i})$ и
    $0 \leqslant t_{0} < t_{1} < t_{2} < \ldots < t_{N} \leqslant T$;
  \item
    $\Delta W_{i} = W(t_{i+1}) - W(t_{i}) \sim
    \mathcal{N}(0,t_{i+1}-t_{i})$,
    где $0\leqslant t_{i+1} < t_{i} < T$, $i=0,1,\ldots,N-1$.
  \end{itemize}
\end{thedefinition}

Обозначение $\Delta W_{i} \sim \mathcal{N}(0,\Delta t_{i})$ говорит о
том, что $\Delta W_{i}$ --- нормально распределенная случайная
величина с математическим ожиданием
$\mathbb{E}[\Delta W_{i}] = \mu = 0$ и дисперсией
$\mathbb{D}[\Delta W_{i}] = \sigma^{2} = \Delta t_{i}$.

Многомерный винеровский процесс
$W^{\alpha}(t)\colon \Omega\times[t_{0},T]\to \mathbb{R}^{m},\;\forall
\alpha = 1,\ldots,m$,
определяют как случайный процесс, составленный из совместно
независимых одномерных винеровских процессов
$W^{1}(t),\ldots,W^{m}(t)$. Приращения $\Delta W^{\alpha}_{i}$
являются совместно независимыми нормально распределенными случайными
величинами.

\subsection{Генерирование выборочных траекторий процесса Винера на
  компьютере}
        
\begin{figure}
  \centering
  \includegraphics[width=0.4\linewidth]{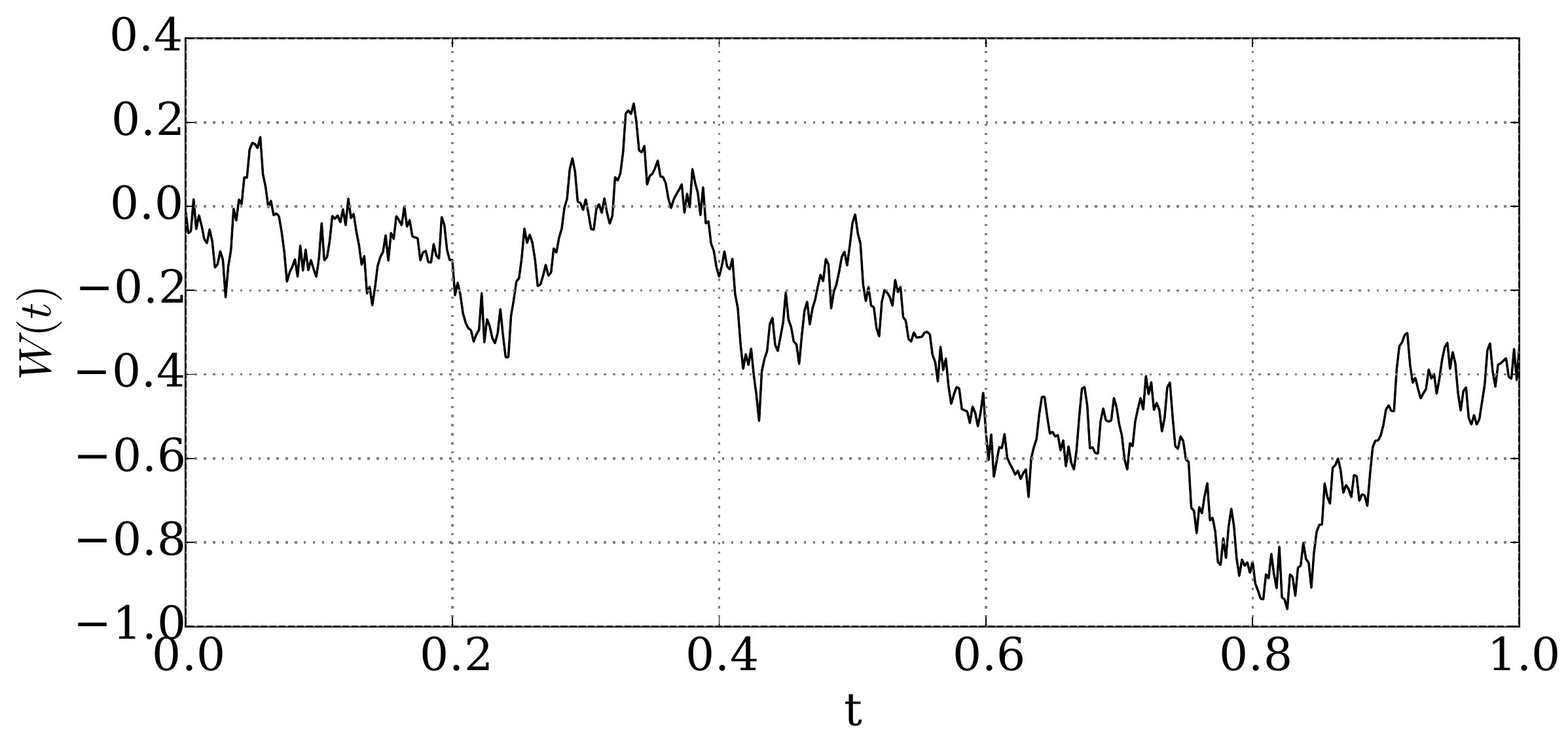}
  \caption{Траектория винеровского процесса}
\label{fig:wiener}
\end{figure}

Для генерирования выборочной траектории одномерного винеровского
процесса на отрезке $[0,T]$, разбитом на $N$ частей с шагом
$\Delta t_{i} = h = \mathrm{const}$, необходимо сгенерировать $N$
нормально распределенных случайных чисел
$\varepsilon_{0}, \ldots, \varepsilon_{N-1} \sim \mathcal{N}(0,h)$ и
построить их накопленные суммы $\varepsilon_{0}$,
$\varepsilon_{0} + \varepsilon_{1}$,
$\varepsilon_{0} + \varepsilon_{1} + \varepsilon_{2}$ и так далее. В
результате получим два массива длины $N$, первый \texttt{W} --- точки
выборочной траектории и второй \texttt{dW} --- приращения этой
траектории. 
Пример траектории приведен на рис.~\ref{fig:wiener}.

В случае $m$-мерного случайного процесса следует сгенерировать уже $m$
последовательностей из $N$ нормально распределенных случайных величин.

В описываемой нами библиотеке для генерирования выборочных
винеровских траекторий написана функция:
\begin{lstlisting}[language=python]
(dt, t, dW, W) = wiener_process(N, dim=1, interval=(0.0, 1.0))
\end{lstlisting}
Обязательны аргумент $N$ --- число точек разбиения временного
интервала \texttt{interval} (по умолчанию $[0,1]$), а \texttt{dim}
--- размерность винеровского процесса. Функция возвращает кортеж из
четырех элементов, где \texttt{dt} --- шаг разбиения
($\Delta t_{i} = h = \mathrm{const}$), \texttt{t} --- одномерный
\texttt{numpy} массив содержащий значения моментов времени
$t_{1},t_{2},\ldots,t_{N}$, \texttt{dW}, \texttt{W} --- также
\texttt{numpy} массивы размерностью $N\times m$ содержащие
приращения $\Delta W^{\alpha}_{i}$ и точки выборочной траектории
$W^{\alpha}_{i}$, где $i=1,\ldots,N;\;\alpha=1,\ldots,m$.

\subsection{СДУ в форме Ито для многомерного процесса Винера}

Введем вероятностное пространство
$(\Omega, \mathscr{A}, \mathbb{P})$, где $\Omega$ --- пространство
элементарных событий, $\mathscr{A}$ --- сигма-алгебра подмножеств
пространства $\Omega$, $\mathbb{P}$ --- вероятностная мера. На
отрезке $[t_{0}, T] \in \setR^{1}$ определена переменная $t$ имеющая
физический смысл времени.

Рассмотрим случайный процесс
$\mathbf{x}(t) = (x^{1}(t), \ldots, x^{d}(t))^{T}$, где
$\mathbf{x}(t)$ принадлежит функциональному пространству
$\mathrm{L}^{2}(\Omega)$ с нормой $\| \cdot \|$. Будем считать, что
случайный процесс $\mathbf{x}(t)$ является решением для СДУ в форме
\emph{Ито}~\cite{Oksendal::ru,Kloeden_Platen}:
\[
x^{\alpha}(t) = f^{\alpha}(t,x^{\gamma}(t))\d t + \sum\limits_{\beta =
  1}^{m}G^{\alpha}_{\beta}(t,x^{\gamma}(t))\d W^{\beta},
\]
где $\alpha,\gamma = 1,\ldots,d$, $\beta = 1,\ldots,m$, функция
$f^{\alpha}(t,x^{\gamma}(t)) = f^{\alpha}(t,x^{1}(t),\ldots,x^{d}(t))$
--- \emph{вектор сноса}, а матричнозначная функция
$g^{\alpha}_{\beta}(t,x^{\gamma}(t))$ --- \emph{матрица диффузии},
$W^{\alpha} = (W^{1},\dots,W^{m})^T$ --- многомерный винеровский
процесс, называемый \emph{ведущим (driver)} процессом СДУ.
Кроме того
$\mathbf{f}(t,\mathbf{x}(t)) =
(f^{1}(t,\mathbf{x}),\ldots,f^{d}(t,\mathbf{x}))^{T}$,
а матрица $\mathbf{G}$ имеет вид:
\[
\mathbf{G} =
\begin{bmatrix}
  g^{1}_{1}(t,\mathbf{x}) & g^{1}_{2}(t,\mathbf{x}) & \ldots &
  g^{1}_{m}(t,\mathbf{x})\\
  g^{2}_{1}(t,\mathbf{x}) & g^{2}_{2}(t,\mathbf{x}) & \ldots &
  g^{2}_{m}(t,\mathbf{x})\\
  \vdots & \vdots & \ddots & \vdots \\
  g^{d}_{1}(t,\mathbf{x}) & g^{d}_{2}(t,\mathbf{x}) & \ldots &
  g^{d}_{m}(t,\mathbf{x})
\end{bmatrix}
\]

На отрезке $[t_{0}, T]$ введем сетку $t_{0}<t_{1}<\ldots<t_{N}=T$ с
шагом $h_{n} = t_{n+1} - t_{n}$, где $n = 0,\ldots,N-1$ и максимальный
шаг сетки $h = \max{\{h_{n-1}\}^{N}_{1}}$. Далее будем полагать, что
если сетка равномерная, то $h_{n} = h = \const$; $\mathbf{x}_{n}$ ---
сеточная функция, аппроксимирующая случайный процесс $\mathbf{x}(t)$,
так что $\mathbf{x}_{0} = \mathbf{x}(t_{0})$,
$\mathbf{x}_{n} \approx \mathbf{x}(t_{n})\;\forall n = 1,\ldots,N$.

\subsection{Слабая и сильная сходимости аппроксимирующей функции}

Необходимо определить критерий точности аппроксимации процесса
$\mathbf{x}(t)$ последовательностью функций
$\{\mathbf{x}_{n}\}^{N}_{1}$. Таких критериев принято выделять два:
\emph{слабый} и \emph{сильный}~\cite{Oksendal::ru,Andreas_2003,Rossler_2010}.

\begin{thedefinition}
  Последовательность аппроксимирующих функций
  $\{\mathbf{x}_{n}\}^{N}_{1}$ сходится в \emph{сильном смысле} с
  порядком $p$ к решению $\mathbf{x}(t)$ СДУ в момент времени $T$,
  если существует константа $C>0$ и $\delta_{0}>0$, такая что
  $\forall h \in (0,\delta_{0}]$ выполняется условие
  \[
  \mathbb{E}(\|\mathbf{x}(T) - \mathbf{x}_{N}\|) \leqslant Ch^{p}.
  \]
\end{thedefinition}

\begin{thedefinition}
  Последовательность аппроксимирующих функций
  $\{\mathbf{x}_{n}\}^{N}_{1}$ сходится в \emph{слабом смысле} с
  порядком $p$ к решению $\mathbf{x}(t)$ СДУ в момент времени $T$,
  если существует константа $C_{F}>0$ и $\delta_{0}>0$, такая что
  $\forall h \in (0,\delta_{0}]$ выполняется условие
  \[
  \left|\mathbb{E}\left[F(\mathbf{x}(T))\right] -
    \mathbb{E}\left[F(\mathbf{x}_{N})\right]\right| \leqslant
  C_{F}h^{p}.
  \]
\end{thedefinition}
Здесь $F \in C^{2(p+1)}_{\mathrm{P}}(\setR,\setR^{d})$ --- непрерывный
дифференцируемый (вплоть до производных $2(p+1)$ порядка) функционал с
полиномиальным ростом.

Если матрица $\mathbf{G}$ обращается в ноль, то условие сильной
сходимости равносильно условию сходимости для детерминированного
случая. В отличие от детерминированного случая порядок сильной
сходимости не обязательно является натуральным числом и может
принимать дробно-рациональные значения.

Выбор типа сходимости целиком зависит от решаемой задачи. Численные
методы с сильной сходимостью хорошо аппроксимируют конкретные
траектории случайного процесса $\mathrm{x}(t)$ и поэтому нуждаются в
информации о траектории ведущего винеровскго процесса. На практике это
означает, что функции, реализующей стохастический метод с сильной
сходимостью, необходимо передать двумерный массив со значениями
приращений $\Delta W^{\alpha}_{n}$, где
$\alpha=1,\ldots,m;\;n=1,\ldots,N$. Этот массив будет использоваться
для аппроксимации кратных стохастических интегралов, которые входят в
любую численную схему порядка $p > \frac{1}{2}$.

В свою очередь слабые численные методы хорошо аппроксимируют
характеристики вероятностного распределения случайного процесса
$\mathrm{x}(t)$. Они не нуждаются в информации о траектории
соответствующего винеровского процесса, и случайные величины для этих
методов могут быть сгенерированы на другом вероятностном пространстве,
которое легко реализовать программно.

\subsection{Вычисление и аппроксимация кратных интегралов Ито}

В общем случае для конструирования численных схем, порядок сильной
сходимости которых был бы больше $p=\frac{1}{2}$, необходимо включение
в формулы этих схем однократных и двукратных интегралов Ито:
\begin{gather}
I^{i}(t_{n}, t_{n+1}) = I^{i}(h_{n}) 
= \int\limits_{t_{n}}^{t_{n+1}}
\mathrm{\d}W^{i}(\tau), \\
I^{ij}(t_{n}, t_{n+1}) = I^{ij}(h_{n}) 
= 
\int\limits_{t_{n}}^{t_{n+1}}\int\limits_{t_{n}}^{\tau_{1}}\mathrm{d}W^{i}(\tau_{2})\mathrm{d}W^{j}(\tau_{1}),
\end{gather}
где $i,j=1\ldots,m$ и $W^{i}$ --- компоненты многомерного винеровского
процесса.

Возникает задача выражения этих интегралов через приращения
$\Delta W^{i}_{n} = W^{i}(t_{n+1}) - W^{i}(t_{n})$. В случае
однократного интеграла это можно сделать для любого индекса $i$:
$I^{i}(h_{n}) = \Delta W^{i}_{n}$. В случае же двукратного интеграла
$I^{ij}(h_{n})$ точная формула имеет место лишь при $i=j$:
\[
I^{ii}(h_{n}) = \dfrac{1}{2}\left( (\Delta W^{i}_{n})^2 - \Delta t_{n}
\right).
\]
В остальных же случаях при $i\neq j$ выразить $I^{ij}(h_{n})$ через
приращения $\Delta W^{\alpha}_{n}$ и $\Delta t_{n}$ в конечном виде не
представляется возможным, поэтому остается лишь использовать численную
аппроксимацию.

В книге~\cite{Kloeden_Platen} приведены следующие формулы для
аппроксимации двукратного интеграла Ито $I^{ij}$:
\begin{gather*}
  I^{ij}(h_{n}) = \dfrac{\Delta W^{i}_{n} \Delta W^{j}_{n} -
  h_{n}\delta^{ij}}{2} + A^{ij}(h_{n}),\\
  A^{ij}(h_{n}) =
  \dfrac{h}{2\pi}\sum\limits^{\infty}_{k=1}\dfrac{1}{k}\left(V^{i}_{k}(U^{j}_{k}
    + \sqrt{2/h_{n}} \Delta W^{j}_{n}) - V^{j}_{k}(U^{i}_{k} +
    \sqrt{2/h_{n}}\Delta W^{i}_{n})\right),
\end{gather*}
$\Delta W^{i}_{n} \sim \mathcal{N}(0,h)$,
$V^{i}_{k} \sim \mathcal{N}(0,1)$,
$U^{i}_{k} \sim \mathcal{N}(0,1)$,
$i=1,\ldots,m;\; k=1,\ldots,\infty$; $i=1,\ldots,N$. Из формул
видно, что в случае $i=j$ получаем конечное выражение для $I^{ii}$,
но в случае $i\neq j$ приходится суммировать бесконечный ряд
$A^{ij}$. Данный алгоритм дает ошибку аппроксимации порядка
$O(h^2/n)$, где $n$ --- число оставленных слагаемых бесконечного
ряда $A^{ij}$.

В статье~\cite{Wiktorsson_2001} предложен матричный вид
аппроксимирующих формул. Пусть $\mathbf{1}_{m\times m}$,
$\mathbf{0}_{m\times m}$ --- единичная и нулевая матрицы $m\times m$.
Тогда
\begin{gather}
  \mathbf{I}(h_{n}) = \dfrac{\Delta \mathbf{W}_{n} \Delta \mathbf{W}_{n} - h_{n}\mathbf{1}_{m\times m}}{2} + \mathbf{A}(h_{n}),\\
  \mathbf{A}(h_{n}) =
  \dfrac{h}{2\pi}\sum\limits^{\infty}_{k=1}\dfrac{1}{k}\left(\mathbf{V}_{k}(\mathbf{U}_{k}
    + \sqrt{2/h_{n}} \Delta \mathbf{W}_{n})^{T} - (\mathbf{U}_{k} +
    \sqrt{2/h_{n}}\Delta \mathbf{W}_{n})\mathbf{V}^{T}_{k}\right),
\end{gather}
где $\Delta \mathbf{W}_{n}, \mathbf{V}_{k}, \mathbf{U}_{k}$ ---
независимые нормально распределенные многомерные случайные величины:
\begin{gather}
  \Delta \mathbf{W}_{n} = (\Delta W^{1}_{n}, \Delta W^{2}_{n}, \ldots, \Delta W^{m}_{n})^{T} \sim \mathcal{N}(\mathbf{0}_{m\times m}, h_{n}\mathbf{1}_{m\times m}),\\
  \mathbf{V}_{k} = (V^{1}_{k}, V^{2}_{k},\ldots, V^{m}_{k})^{T} \sim
  \mathcal{N}(\mathbf{0}_{m\times m}, \mathbf{1}_{m\times m}),\;\;
  \mathbf{U}_{k} = (U^{1}_{k}, U^{2}_{k},\ldots, U^{m}_{k})^{T} \sim
  \mathcal{N}(\mathbf{0}_{m\times m}, \mathbf{1}_{m\times m}).
\end{gather}

В нашей библиотеке написан ряд функций для аппроксимации интегралов
Ито. Следующие две функции предназначены для вычисления интегралов
Ито кратности 1 и 2 для одномерного процесса Винера:
\begin{lstlisting}[language=python]
I1 = Ito1W1(dW); I2 = Ito2W1(dW, h)
\end{lstlisting}
они возвращают \texttt{numpy} массивы разной размерности. Функция
\texttt{Ito1W1} тривиальна и возвращает массив приращений
винеровского процесса, в функция \texttt{Ito2W1} возвращает
\texttt{numpy} массив размерности \texttt{(N, 2, 2)}, что
соответствует интегралам
$I^{ij}(h_{n})\;\forall n = 1\ldots N,\;i,j=0,1$.

Следующие две функции предназначены для вычисления аппроксимации
интегралов Ито кратности 1 и 2 для многомерного процесса Винера:
\begin{lstlisting}[language=python]
I1m = Ito1Wm(dW, h); I2m = Ito2Wm(dW, h, n).
\end{lstlisting}
Они возвращают \texttt{numpy} массивы разной размерности. Функция
\texttt{Ito1Wm} проводит вычисления на основе точных формул и
возвращает свой массив размерности \texttt{(N, m+1)}, а функция
\texttt{Ito2Wm} возвращает \texttt{numpy} массив размерности
\texttt{(N, m+1, m+1)}, что соответствует интегралам
$I^{ij}(h_{n})\;\forall n = 1\ldots N,\;i,j=0,\ldots,m$. Аргумент
\texttt{n} задает число членов в бесконечном ряде $A^{ij}$, который
используется для аппроксимации стохастического интеграла.

\subsection{СДУ имеющие аналитическое решение}

Для верификации написанных программ в случае сильной сходимости будем
использовать СДУ с известным аналитическим решением, что даст
возможность вычислить погрешности аппроксимации реализаций траекторий
ведущих винеровских процессов.

\subsubsection{Логарифмическое блуждание (одномерный винеровский
  процесс)}

В качестве скалярного СДУ с известным аналитическим решением возьмем
уравнение \emph{логарифмического
  блуждания}~\cite{Platen_Liberati_2010,Kloeden_Platen}:
\[
\mathrm{d}x(t) = \mu x(t) \mathrm{d}t + \sigma x(t)
\mathrm{d}W(t),\;\; x(0) = x_{0},
\]
точное решение которого имеет вид~\cite[пункт 4.4]{Kloeden_Platen}:
\[
x(t) = x_0 \exp{\left((\mu - 0.5\sigma^2)t + \sigma W(t)\right)}.
\]
При вычислениях будем использовать $\mu = 2,\; \sigma = 1$ и
$ x_{0} = 1$.

\subsubsection{Двумерная модель Блека--Шоулза}

Для тестирования численных методов с сильной сходимостью для систем
СДУ с многомерным винеровским процессом возьмем уравнение двумерной
модели Блека--Шоулза~\cite{Platen_Liberati_2010}, которое обычно
записывается в следующей форме:
\[
\mathbf{x}(t) = \mathrm{A}\mathbf{x}(t)\mathrm{d}t +
\sum\limits^{2}_{k=1}\mathrm{B}_k \mathbf{x}(t)\mathrm{d}W^{k},
\]
где все матрицы диагональны: $A = \mathrm{diag}(a^1,a^2)$,
$B_{1} = \mathrm{diag}(b^1_1,b^2_1)$ и
$B_{2} = \mathrm{diag}(b^1_2,b^2_2)$. После ряда преобразования можно
выразить функции $\mathbf{f}(\mathbf{x})$ и $\mathrm{G}(\mathbf{x})$ в
явном виде
\[
\mathrm{f}(\mathbf{x}) =
\begin{bmatrix}
  a^{1}x^{1}(t)\\
  a^{2}x^{2}(t)
\end{bmatrix},\;\; \mathrm{G}(\mathbf{x}) =
\begin{bmatrix}
  b^1_1 x^{1}(t)&b^1_2 x^{1}(t)\\
  b^2_1 x^{2}(t)&b^2_2 x^{2}(t)
\end{bmatrix}.
\]
Аналитическое решение можно записать в виде экспоненты, которая
берется поэлементно
\[
\mathbf{x}(t) = \mathbf{x}_0 \exp \left\{\left(\mathrm{A} -
    \frac{1}{2}\sum\limits^{2}_{k=1}B_k^2\right)t +
  \sum\limits^{2}_{k=1}\mathrm{B}_k W^k (t)\right\}.
\]
Для проведения вычислений возьмем следующие матрицы:
$A = \mathrm{diag}(a_1,a_2)$, $B_{1} = \mathrm{diag}(b_1,b_2\rho)$ и
$B_{2} = \mathrm{diag}(0,b_2\sqrt{1-\rho})$
и следующие значения параметров $x^{1}_{0} = x^{1}_{0} = 1.0$,
$a_{1} = a_{2} = 0.1$, $b_{1} = b_{2} = 0.2$, $\rho = 0.8$. Тогда
точное решение уравнения примет вид:
\begin{align*}
  &x^{1}(t) = x^{1}_{0}\exp\left((a_{1} - 0.5b^{2}_{1})t + b_{1}W^{1}(t)\right)\\
  &x^{2}(t) = x^{2}_{0}\exp\left((a_{2} - 0.5b^{2}_{2})t + b_{2}\left(\rho W^{1}(t) + \sqrt{1+\rho^2}W^{2}(t)\right)\right).
\end{align*}

\section{Стохастические методы Рунге--Кутты}

В данном разделе мы приведем несколько наиболее эффективных
стохастических численных схем Рунге--Кутты, основываясь на результатах
работ~\cite{Rossler_2010,Andreas_2003,Debrabant_2007,Debrabant_2013,Tocino_2001,Soheili_2007,Mackevicius_1994}. Ограничимся
численными схемами без производных от
вектора сноса и матрицы диффузии, поэтому эффективные схемы Мильштейна
останутся вне нашего
внимания~\cite{Milstein_1974,Milstein_1979,Milstein_1986}.

Вначале укажем ряд факторов, которые делают стохастические методы
Рунге-Кутты существенно сложнее классических методов:
\begin{itemize}
\item При выборе конкретного метода надо учитывать, какой тип
сходимости необходимо обеспечить для данной конкретной задачи, а
также какое из стохастических уравнений необходимо решать --- в
форме Ито или в форме Стратоновича. Это увеличивает количество
алгоритмов, которые нужно запрограммировать.
\item Для методов с сильной сходимостью $p_{s} > 1$. На каждом шаге
необходимо решать ресурсоемкую задачу по аппроксимации двукратных
стохастических интегралов.
\item В численной схеме присутствуют не только матрицы и векторы, но и
тензоры (четырехмерные массивы), с которыми необходимо совершать
операцию свертки по нескольким индексам. Реализация свертки через
суммирование с помощью обычных циклов приводит к существенному
падению производительности.
\item Для использования слабых методов необходимо применять метод
Монте-Карло, проводя несколько серий по $10^7$--$10^8$ испытаний в
каждой.
\end{itemize}

Наиболее существенно на скорости работы программы сказывается
необходимость свертки многомерных массивов. Для ускорения этой
операции при реализации численных схем была использована функция
\texttt{einsum} из библиотеки \texttt{numpy} версии 1.6 и старше. В
случае использования обычных последовательных циклов
производительность вычислений существенно падает.
\begin{lstlisting}[language=python]
numpy.einsum(subscripts, *operands, out=None, dtype=None, order='K', casting='safe')
\end{lstlisting}

Также следует обратить внимание на большое количество нулей в
обобщенных таблицах Бутчера. Реализация отдельной функции для
конкретной реализации метода часто помогает получить выигрыш в
производительности в несколько раз по сравнению с функцией, которая
реализует универсальный алгоритм.

\subsection{Численный метод Эйлера--Маруйамы}

Простейшим численным методом для решения как скалярных уравнений, так
и систем СДУ, является метод Эйлера--Маруямы, названный так в честь
Гиширо Маруямы (Gisiro Maruyama), который распространил классический
метод Эйлера для ОДУ на случай СДУ~\cite{Maruyama_1955}. Метод легко
обобщается на случай многомерного винеровского процесса:
\begin{align*}
&x^{\alpha}_{0} = x^{\alpha}(t_{0}),\\
&x^{\alpha}_{n+1} = x^{\alpha}_{n} + f^{\alpha}(t_{n},
x^{\alpha}_{n})h_{n} +
\sum\limits^{d}_{\gamma=1}G^{\alpha}_{\beta}(t_{n},x^{\gamma}_{n})\Delta
W^{\beta}_{n}.
\end{align*}

Как видно из формул, на каждом шаге для вычисления следующего значения
аппроксимирующей функции требуется лишь соответствующее данному шагу
приращение процесса $\Delta W^{\beta}_{n}$. Метод имеет сильный
порядок $(p_{d},p_{s}) = (1.0, 0.5)$. Величина $p_{d}$ обозначает
порядок точности детерминированной части численного метода, то есть
той точности, которую будет давать численный метод при применении его
к уравнению функции $G(t,x^{\alpha}(t))\equiv 0$. Величина $p_{s}$
обозначает порядок приближения стохастической части уравнения.

\subsection{Стохастические методы Рунге--Кутты сильной сходимости
$p=1.5$ для одномерного винеровского процесса}

В случае скалярных уравнения и винеровского процесса справедлива
следующая численная схема:
\begin{align}
&X^{i}_{0} = x_{n} + \sum\limits^{s}_{j=1}A_{0j}^{i}f(t_{n} + c^{j}_{0}h_{n}, X^{j}_{0})h_{n} + \sum\limits^{s}_{j=1}B^{i}_{0j}g(t_{n} + c^{j}_{1}h_{n}, X^{j}_{1})\dfrac{I^{10}(h_{n})}{\sqrt{h_{n}}},\\
&X^{i}_{1} = x_{n} + \sum\limits^{s}_{j=1}A_{1j}^{i}f(t_{n} + c^{j}_{0}h_{n}, X^{j}_{0})h_{n} + \sum\limits^{s}_{j=1}B^{i}_{1j}g(t_{n} + c^{j}_{1}h_{n}, X^{j}_{1})\sqrt{h_{n}},\\
&x_{n+1} = x_{n} + \sum\limits^{s}_{i=1}a_{i}f(t_{n}+c^{i}h_{n}, X^{i}_{0})h_{n} + \\
& \qquad + \sum\limits^{s}_{i=1}\left(b^{1}_{i}I^{1}(h_{n}) + b^{2}_{i}\dfrac{I^{11}(h_{n})}{\sqrt{h_{n}}} + b^{3}_{i}\dfrac{I^{10}(h_{n})}{h_{n}} + b^{4}_{i}\dfrac{I^{111}(h_{n})}{h_{n}}\right)g(t_{n} + c^{i}_{1},X^{i}_{1}).
\end{align}
Обобщённая таблица Бутчера который имеет вид~\cite{Rossler_2010}:
\[ {\renewcommand{\arraystretch}{1.5}%
\begin{array}{c|c|c|c}
c^{i}_{0} & A^{i}_{0j} & B^{i}_{0j} &\\ \hline
c^{i}_{1} & A^{i}_{1j} & B^{i}_{1j} &\\ \hline
                                        & a_{i} & b^{1}_{i} & b^{2}_{i}\\ \hline
                                        &  & b^{3}_{i} & b^{4}_{i}\\
\end{array}}.
\]

В препринте Росслера~\cite{Rossler_2010} для данной схемы приведены
две таблицы Бутчера для метода четвертой стадийности $s=4$:
{\scriptsize%
\[
{\renewcommand{\arraystretch}{1.5}%
\begin{array}{c|cccc|cccc|cccc}
        0&0&0&0&0&0&0&0&0&&&\\
        \frac34&\frac34&0&0&0&\frac{3}{2}&0&0&0&&&\\
        0&0&0&0&0&0&0&0&0&&&\\
        0&0&0&0&0&0&0&0&0&&&\\ \hline
        0&0&0&0&0&0&0&0&0&&&\\
        \frac14&\frac14&0&0&0&\frac12&0&0&0&&&\\
        1&1&0&0&0&-1&0&0&0&&&\\
        \frac14&0&0&\frac14&0&-5&3&\frac12&0&&&\\ \hline
        &\frac13&\frac23&0&0&-1&\frac43&\frac23&0&-1&\frac43&-\frac13&0\\ \hline
        &&&&&2&-\frac43&-\frac23&0&-2&\frac53&-\frac23&1
\end{array}}\;\;
{\renewcommand{\arraystretch}{1.5}%
\begin{array}{c|cccc|cccc|cccc}
        0&0&0&0&0&0&0&0&0&&&\\
        1&\frac34&0&0&0&0&0&0&0&&&\\
        \frac12&\frac14&\frac14&0&0&1&\frac12&0&0&&&\\
        0&0&0&0&0&0&0&0&0&&&\\ \hline
        0&0&0&0&0&0&0&0&0&&&\\
        \frac14&\frac14&0&0&0&-\frac12&0&0&0&&&\\
        1&1&0&0&0&1&0&0&0&&&\\
        \frac14&0&0&\frac14&0&2&-1&\frac12&0&&&\\ \hline
        &\frac16&\frac16&\frac23&0&-1&\frac43&\frac23&0&1&-\frac43&\frac13&0\\ \hline
        &&&&&2&-\frac43&-\frac23&0&-2&\frac53&-\frac23&1
\end{array}}.
\]}

Численную схему, реализуемую первой таблицей, обозначим как
\texttt{SRK1W1}, а вторую --- как \texttt{SRK2W2}. Метод
\texttt{SRK1W1} имеет сильный порядок $(p_{d},p_{s}) = (2.0, 1.5)$, а
метод \texttt{SRK2W1} --- сильный порядок
$(p_{d},p_{s}) = (3.0, 1.5)$. Еще один метод с сильным порядком
$p_{s}=1.0$ приведен в книге~\cite{Kloeden_Platen} и его таблица
Бутчера имеет вид:
{\scriptsize%
\[
\text{\texttt{KlPl}: } {\renewcommand{\arraystretch}{1.1}%
\begin{array}{c|cc|cc|cc}
        0&0&0&0&0&&\\
        0&0&0&0&0&&\\ \hline
        0&0&0&0&0&&\\
        0&0&0&0&0&&\\ \hline
        0&1&0&1&0&-1&1\\
        & & &1&0&0&0\\
\end{array}}
\]}

На рис.~\ref{fig:fig1} приведены графики локальной погрешности
аппроксимации точного решения стохастического уравнения модели
логарифмического блуждания. Для решения использовались реализации
стохастического метода Рунге-Кутты сильной сходимости $p_s = 1.5$.
Отметим, что разный порядок аппроксимации детерминированной части
уравнения в методах \texttt{SRK1W1} и \texttt{SRK2W1} несущественно
влияет на общую погрешность вычисления.

\begin{minipage}[b]{0.45\linewidth}
\begin{figure}[H]
\centering
\includegraphics[width=1.0\linewidth]{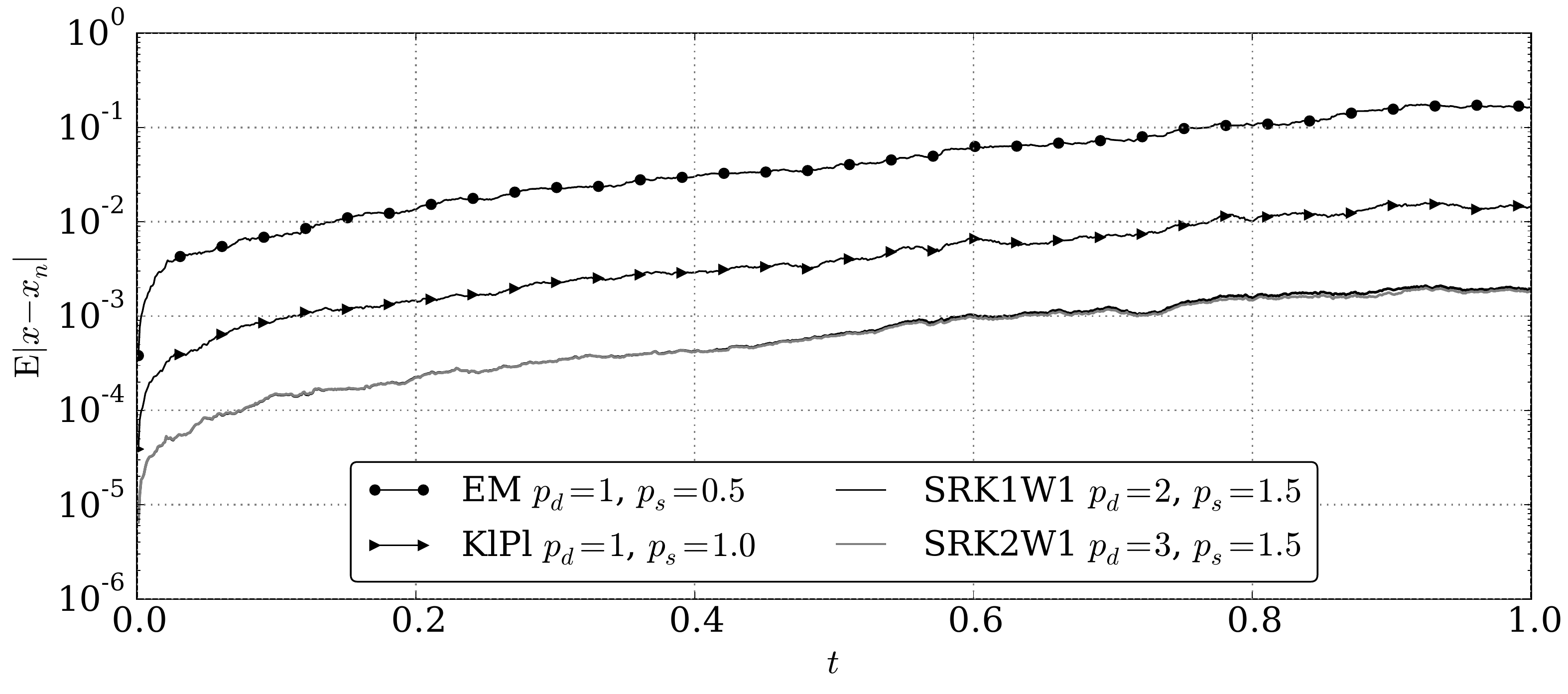}
\caption{Сильные погрешности методов для скалярного процесса $W$ c
        $h=10^{-3}$}
\label{fig:fig1}
\end{figure}
\end{minipage}
\hfill
\begin{minipage}[b]{0.45\linewidth}
\begin{figure}[H]
\centering
\includegraphics[width=1.0\linewidth]{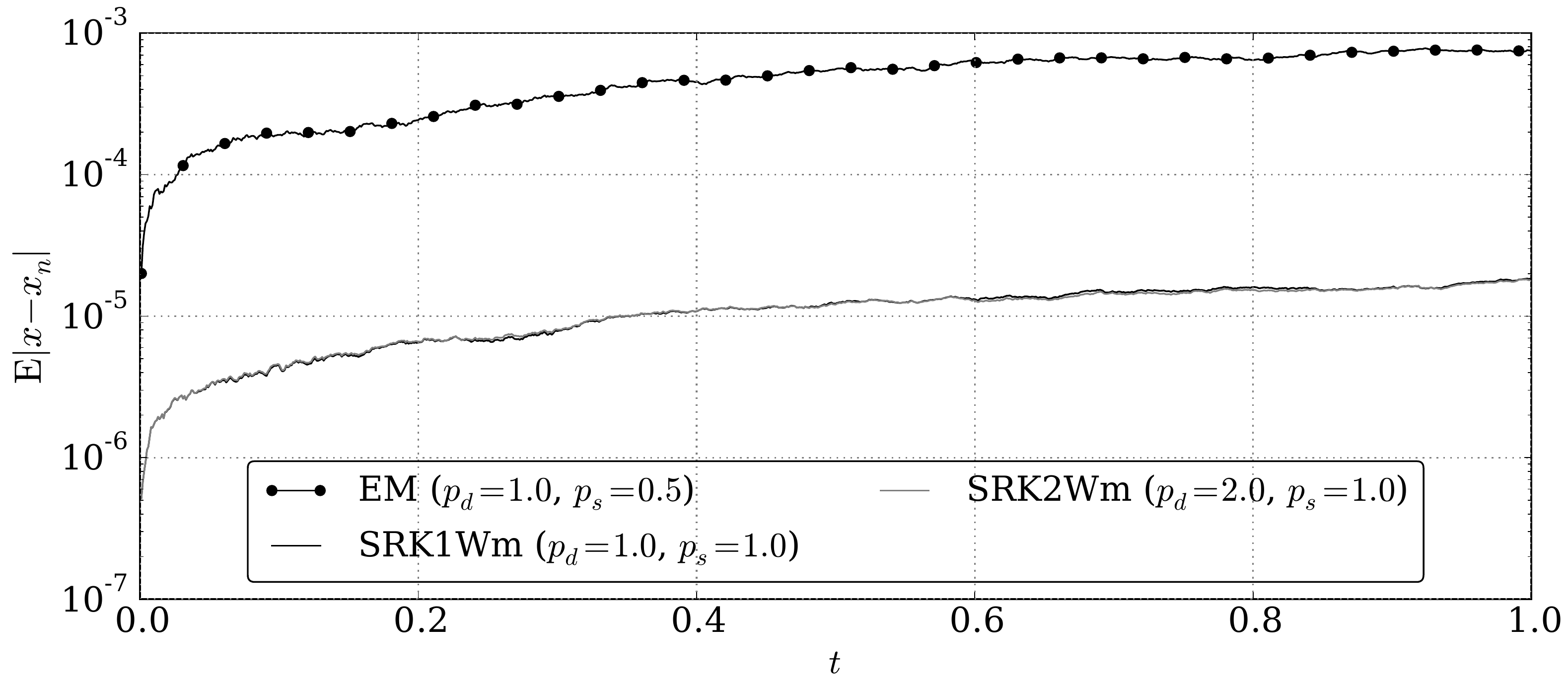}
\caption{Сильные погрешности методов для двумерного процесса
        $W^{\alpha}$ c $h=10^{-3}$}
\label{fig:fig2}
\end{figure}
\end{minipage}

В описываемой библиотеке написано несколько функций для решения СДУ
со скалярным винеровским процессом
\begin{lstlisting}[language=python]
x_num = EulerMaruyama(f, g, h, x_0, dW) x_num = strongSRKW1(f, g,
h,
x_0, dW, name='SRK2W1')
\end{lstlisting}
где \texttt{f(x)} и \texttt{g(x)} --- функции, принимающие в
качестве аргумента и возвращающие действительное число; \texttt{h}
--- шаг сетки, \texttt{x\_0} --- начальное значение, \texttt{dW} ---
массив приращений винеровского процесса \texttt{N}; аргумент
\texttt{name} может принимать значения \texttt{'SRK1W1'} и
\texttt{'SRK2W2'}.

\subsection{Стохастические методы Рунге--Кутты сильной сходимости $p=1.0$ 
для многомерного винеровского потока}

Для системы СДУ Ито с многомерным винеровским процессом можно
построить стохастическую численную схему Рунге-Кутты сильного порядка
$p_{s} = 1.0$ с использованием однократных и двукратных интегралов
Ито~\cite{Rossler_2010}.

\begin{align*}
&X^{0i\alpha} = x^{\alpha}_{n} + \sum\limits^{s}_{j=1}A^{i}_{0j}f^{\alpha}(t_{n} + c^{j}_{0}h_{n}, X^{0j\beta})h_{n} + \sum\limits_{l=1}^{m}\sum\limits^{s}_{j=1}B^{i}_{0j}G^{\alpha}_{l}(t_{n} + c^{j}_{1}h_{n},X^{lj\beta})I^{l}(h_{n}),\\
&X^{ki\alpha} = x^{\alpha}_{n} + \sum\limits^{s}_{j=1}A^{i}_{1j}f^{\alpha}(t_{n} + c^{j}_{0}h_{n},X^{0j\beta})h_{n} + \sum\limits_{l=1}^{m}\sum\limits^{s}_{j=1}B^{i}_{1j}G^{\alpha}_{l}(t_{n} + c^{j}_{1}h_{n},X^{lj\beta})\dfrac{I^{lk}(h_{n})}{\sqrt{h_{n}}},\\
&x^{\alpha}_{n+1} = x^{\alpha}_{n} + \sum\limits^{s}_{i=1}a_{i}f^{\alpha}(t_{n} + c^{i}_{0}h_{n},X^{0i\beta})h_{n} + \sum\limits^{m}_{k=1}\sum\limits^{s}_{i=1}(b^1_{i}I^{k}(h_{n}) + b^2_{i}\sqrt{h_{n}})G^{\alpha}_{k}(t_{n} + c^{i}_{1}h_{n},X^{ki\beta}),
\end{align*}
где $n=0,1,\ldots,N-1$; $i=1,\ldots,s$; $\beta,k=1,\ldots,m$;
$\alpha = 1,\ldots,d$. 

Обобщённый таблица Бутчера имеет
вид~\cite{Rossler_2010}:
\[ {\renewcommand{\arraystretch}{1.4}%
\begin{array}{c|c|c|c}
c^{i}_0 & A^{i}_{0j} & B^{i}_{0j} &\\ \hline
c^{i}_1 & A^{i}_{1j} & B^{i}_{1j} &\\ \hline
                                & a_{i} & b^{1}_{i} & b^{2}_{i} \\
\end{array}}.
\]

В препринте Росслера~\cite{Rossler_2010} указаны две таблицы Бутчера
для метода третей стадийности $s=3$:
{\scriptsize%
\[
\text{\texttt{SRK1Wm}: } {\renewcommand{\arraystretch}{1.1}%
\begin{array}{c|ccc|ccc|ccc}
        0&0&0&0&0&0&0&&&\\
        0&0&0&0&0&0&0&&&\\
        0&0&0&0&0&0&0&&&\\ \hline
        0&0&0&0&0&0&0&&&\\
        0&0&0&0&1&0&0&&&\\
        0&0&0&0&-1&0&0&&&\\ \hline
        &1&0&0&1&0&0&0&\frac{1}{2}&\frac{1}{2}
\end{array}}\;\;
\text{\texttt{SRK2Wm}: }
{\renewcommand{\arraystretch}{1.1}%
\begin{array}{c|ccc|ccc|ccc}
        0&0&0&0&0&0&0&&&\\
        1&1&0&0&0&0&0&&&\\
        0&0&0&0&0&0&0&&&\\ \hline
        0&0&0&0&0&0&0&&&\\
        1&1&0&0&1&0&0&&&\\
        1&1&0&0&-1&0&0&&&\\ \hline
        &\frac{1}{2}&\frac{1}{2}&0&1&0&0&0&\frac{1}{2}&-\frac{1}{2}
\end{array}}
\]}

Метод \texttt{SRK1Wm} имеет сильный порядок сходимости
$(p_{d},p_{s}) = (1.0, 1.0)$, а метод \texttt{SRK2Wm} --- сильный
порядок сходимости $(p_{d},p_{s}) = (2.0, 1.0)$.

На рис.~\ref{fig:fig2} приведены графики локальной погрешности
аппроксимации точного решения двумерной модели Блека--Шоулза. Для
решения использовались реализации стохастического метода Рунге--Кутты
сильной сходимости $p_s = 1.0$ для многомерного винеровского
процесса. Отметим, что и здесь разный порядок аппроксимации
детерминированной части уравнения в методах \texttt{SRK1Wm} и
\texttt{SRK2Wm} несущественно сказывается на общей погрешности
вычисления.

В описываемой библиотеке для решения систем СДУ с многомерным
процессом Винера (предполагается, что $d=m$) написано несколько
функций
\begin{lstlisting}[language=python]
x_num = strongSRKp1Wm(f, G, h, x_0, dW, name='SRK1Wm') x_num =
EulerMaruyamaWm(f, G, h, x_0, dW)
\end{lstlisting}
где \texttt{f(x)} и \texttt{G(x)} --- функции, принимающие в
качестве аргумента массив \texttt{x} и возвращающие numpy-массивы
размерности \texttt{m} и \texttt{(m,m)}; \texttt{h} --- шаг сетки,
\texttt{x\_0} --- начальное значение (одномерный массив размерности
$m$), \texttt{dW} --- массив приращений винеровского процесса
\texttt{(N,m)}; аргумент \texttt{name} может принимать значения
\texttt{'SRK1Wm'} и \texttt{'SRK2Wm'}.

\subsection{Стохастические методы Рунге--Кутты слабой сходимости $p=2.0$}

Численные методы со слабой сходимостью хорошо аппроксимируют
характеристики распределения случайного процесса
$x^{\alpha}(t)$. Слабый численный метод не нуждается в информации о
траектории винеровского процесса $W^{\alpha}_{n}$, и случайные
величины для этих методов могут быть сгенерированы на другом
вероятностном пространстве. Поэтому можно использовать легко
моделируемое на компьютере распределение.

\begin{align*}
& X^{0i\alpha} = x^{\alpha}_{n} + \sum\limits^{s}_{j=1}A_{0j}^{i}f^{\alpha}(t_{n} + c^{j}_{0}h_{n}, X^{0j\beta})h_{n} + \sum\limits^{s}_{j=1}\sum\limits^{m}_{l=1}B_{0j}^{i}G^{\alpha}_{l}(t_{n}+c^{j}_{1}h_{n}, X^{lj\beta})\hat{I}^{l},\\
& X^{ki\alpha} = x^{\alpha}_{n} + \sum\limits^{s}_{j=1}A^{i}_{1j}f^{\alpha}(t_{n}+c^{j}_{0}h_{n}, X^{0j\beta})h_{n} + \sum\limits^{s}_{j=1}B^{i}_{1j}G^{\alpha}_{k}(t_{n} + c_{1}^{j}h_{n}, X^{kj\beta})\sqrt{h_{n}},\\
& \widehat{X}^{ki\alpha} = x^{\alpha}_{n} + \sum\limits^{s}_{j=1}A^{i}_{2j}f^{\alpha}(t_{n}+c^{j}_{0}h_{n}, X^{0j\beta})h_{n} + \sum\limits^{s}_{j=1}\sum\limits^{m}_{l=1,l\neq k}B^{i}_{2j}G^{\alpha}_{l}(t_{n} + c_{1}^{j}h_{n}, X^{lj\beta})\frac{\hat{I}^{kl}}{\sqrt{h_{n}}},\\
& x^{\alpha}_{n+1} = x^{\alpha}_{n} + \sum\limits^{s}_{i=1}a_{i}f^{\alpha}(t_{n}+c^{i}_{1}, X^{ki\beta})h_{n} + \sum\limits^{s}_{i=1}\sum\limits^{m}_{k=1}\left(b^{1}_{i}\hat{I}^{k} + b^{2}_{i}\frac{\hat{I}^{kk}}{\sqrt{h_{n}}}\right) G^{\alpha}_{k}(t_{n}+c^{i}_{1}h_{n}, X^{ki\beta}) + \\
& +  \sum\limits^{s}_{i=1}\sum\limits^{m}_{k=1}\left(b^{3}_{i}\hat{I}^{k} + b^{4}_{i}\sqrt{h_{n}}\right)G^{\alpha}_{k}(t_{n}+c^{i}_{2}h_{n}, \widehat{X}^{ki\beta}).
\end{align*}

Обобщенная таблица Бутчера в этом случае имеет
вид~\cite{Rossler_2010}:
\[
{\renewcommand{\arraystretch}{1.5}%
\begin{array}{c|c|c|c}
c^{i}_{0} & A^{i}_{0j} & B^{i}_{0j} &\\ \hline
c^{i}_{1} & A^{i}_{1j} & B^{i}_{1j} &\\ \hline
c^{i}_{2} & A^{i}_{2j} & B^{i}_{2j} &\\ \hline
                                        & a_{i} & b^{1}_{i} & b^{2}_{i}\\ \hline
                                        &  & b^{3}_{i} & b^{4}_{i}\\
\end{array}}
\] {\scriptsize%
\[
{\renewcommand{\arraystretch}{1.5}%
\begin{array}{c|ccc|ccc|ccc}
        0&0&0&0&0&0&0&&&\\
        1&1&0&0&\frac13&0&0&&&\\
        \frac{5}{12}&\frac{25}{144}&\frac{35}{144}&0&-\frac56&0&0&&&\\ \hline
        0&0&0&0&0&0&0&&&\\
        \frac14&\frac14&0&0&\frac12&0&0&&&\\
        \frac14&\frac14&0&0&-\frac12&0&0&&&\\ \hline
        0&0&0&0&0&0&0&&&\\
        0&0&0&0&1&0&0&&&\\
        0&0&0&0&-1&0&0&&&\\ \hline
        &\frac{1}{10}&\frac{3}{14}&\frac{24}{35}&1&-1&-1&0&1&-1\\ \hline
        &&&&\frac12&-\frac14&-\frac14&0&\frac12&-\frac12\\
\end{array}}\quad
{\renewcommand{\arraystretch}{1.5}%
\begin{array}{c|ccc|ccc|ccc}
        0&0&0&0&0&0&0&&&\\
        1&1&0&0&1&0&0&&&\\
        0&0&0&0&0&0&0&&&\\ \hline
        0&0&0&0&0&0&0&&&\\
        1&1&0&0&1&0&0&&&\\
        1&1&0&0&-1&0&0&&&\\ \hline
        0&0&0&0&0&0&0&&&\\
        0&0&0&0&1&0&0&&&\\
        0&0&0&0&-1&0&0&&&\\ \hline
        &\frac{1}{2}&\frac{1}{2}&0&\frac12&\frac14&\frac14&0&\frac12&-\frac12\\ \hline
        &&&&-\frac12&\frac14&\frac14&0&\frac12&-\frac12\\
\end{array}}.
\]}

В численной схеме слабого стахостического метода Рунге--Кутты
используются следующие случайные величины:
\[
\hat{I}^{kl} = \left\{
\begin{aligned}
&\dfrac{1}{2}(\hat{I}^{k}\hat{I}^{l} - \sqrt{h}_{n}\tilde{I}^{k}),\; k<l,\\
&\dfrac{1}{2}(\hat{I}^{k}\hat{I}^{l} + \sqrt{h}_{n}\tilde{I}^{l}),\; l<k,\\
&\dfrac{1}{2}((\hat{I}^{k})^2 - h_{n}).\; k=l.
\end{aligned}
\right.
\]
Величина $\hat{I}^{k}$ имеет трехточечное распределение. Это означает,
что $\hat{I}^{k}$ может принимать три значения
$ \{-\sqrt{3h_{n}}, 0, \sqrt{3h_{n}}\}$ с вероятностями $1/6$, $2/3$ и
$1/6$ соответственно. В свою очередь величина $\tilde{I}^{k}$ имеет
двухточечное распределение $\{-\sqrt{h_{n}}, \sqrt{h_{n}}\}$ с
вероятностями $1/2$ и $1/2$.

В описываемой библиотеке для генерирования массива случайных чисел,
распределенных по $n$--точечному распределению, написана функция
\begin{lstlisting}[language=python]
n_point_distribution(values, probabilities, shape)
\end{lstlisting}
где аргумент \texttt{values} задает список или массив из $n$ точек
$x_{1},\ldots,x_{n}$, аргумент \texttt{probabilities} задает
массив/список вероятностей $p_{1},\ldots,p_{n}$, а аргумент
\texttt{shape} форму требуемого массива генерируемых случайных
чисел. Для генерирования величин $\hat{I}^{kl}$ предназначена
функция
\begin{lstlisting}[language=python]
(I_hat, I) = weakIto(h, dim, N)
\end{lstlisting}
где \texttt{h} --- шаг сетки, \texttt{dim} --- размерность
винеровского процесса и \texttt{N} --- число точек разбиения отрезка
интегрирования. Возвращаемые величины $\hat{I}^{k}$ и
$\hat{I}^{kl}$.

\subsection{Оценка сильной погрешности}

Для оценки сильной сходимости достаточно сгенерировать одну траекторию
ведущего процесса Винера и для этой траектории рассчитать точное
решение с помощью аналитической формулы, а также приближенное решения
с помощью численных методов. После этого можно найти как локальную
$\|\mathbf{x}(t_{n}) - \mathbf{x}_{n}\|$, $n=1,\ldots,N$, так и
глобальную $\|\mathbf{x}(T) - \mathbf{x}_{N}\|$ ошибки аппроксимации.

\subsection{Оценка слабой погрешности}

Оценка слабой сходимости --- значительно более трудоемкая задача, для
решения которой используется метод Монте-Карло.

Проводится $M$ испытаний, то есть к исходному СДУ $M$ раз применяется
изучаемый численный метод. В результате получаем множество независимых
одинаково распределенных случайных величин $\mathbf{x}_{nm}$, где
$m=1,\ldots,M$ --- номер испытания методом Монте-Карло, а
$n=1,\ldots,N$ --- номер шага метода. Так как обычно интересуются
глобальной погрешностью метода, то рассматривают значения
аппроксимирующей функции на конце отрезка интегрирования $[t_{0}, T]$,
где $t_{N}=T$, или же в любой фиксированной промежуточной точке
отрезка.

Необходимо аппроксимировать функционалы вида
$\mathbb{E}[f(\mathbf{x}(t))]$ с помощью выборочного среднего
(эмпирического математического ожидания) от
$\{\mathbf{x}_{m}\}^{M}_{1}$:
\[
\bar{f}(\mathbf{x}_{N}) =
\frac{1}{M}\sum\limits^{M}_{m=1}f(\mathbf{x}_{m}).
\]

Метод Монте-Карло основывается на центральной предельной теореме,
которая позволяет дать оценку абсолютного значения отклонения
выборочного среднего от математического ожидания величины
$\mathbf{x}$:
\[
\frac{1}{M}\sum\limits^{M}_{m=1}f(\mathbf{x}_{m}) -
\mathbb{E}[f(\mathbf{x})] \xrightarrow[M\to\infty]{}
\mathcal{N}\left(0, \frac{\mathbb{D}[f(\mathbf{x})]}{M}\right)
\]
или иначе
\[
\mathbb{D}\left[\bar{f}(\mathbf{x}) - \mathbb{E}[f(\mathbf{x})]\right]
= \frac{\mathbb{D}[f(\mathbf{x})]}{M}
\]
\[
\mathbb{D}\left[\bar{f}(\mathbf{x}) - \mathbb{E}[f(x)]\right] =
\mathbb{E}\left[\left(\bar{f}(\mathbf{x}) - \mathbb{E}[f(x)] -
0\right)^2\right] = \mathbb{E}\left[\bar{f}(\mathbf{x}) -
\mathbb{E}[f(x)]\right]^2.
\]

Так как выполняется неравенство
$\mathbb{D}[f(\mathbf{x})] = \mathbb{E}[f(\mathbf{x})]^{2} -
(\mathbb{E}[f(\mathbf{x})])^2 \leqslant \mathbb{E}[f(\mathbf{x})]^2$,
то справедлива следующая оценка среднеквадратичной сходимости метода
Монте-Карло:
\[
\sqrt{\mathbb{E}\left[\bar{f}(\mathbf{x}) - \mathbb{E}[f(x)]\right]^2}
\leqslant \frac{\sqrt{\mathbb{E}[f(\mathbf{x})]^2}}{\sqrt{M}}.
\]

Таким образом, Метод Монте-Карло сходится как $1/\sqrt{M}$, и, кроме
того, на его сходимость влияет значение второго момента функции
$f(\mathbf{x})$. Если это значение велико, то число испытаний методом
Монте-Карло необходимо увеличить.

Чтобы оценить глобальную слабую погрешность вычислений, необходимо
учесть как погрешность метода Монте-Карло, так и погрешность самой
численной схемы:
\[
\mathrm{err} = \left| \mathbb{E}[f(\mathbf{x}(T))] -
\frac{1}{M}\sum\limits^{M}_{m=1}f(\mathbf{x}_{Nm})\right| \leqslant
\left| \mathbb{E}[f(\mathbf{x}(T))] -
\mathbb{E}[f(\mathbf{x}_{N})]\right| + \left|
\mathbb{E}[f(\mathbf{x_{N}})] -
\frac{1}{M}\sum\limits^{M}_{m=1}f(\mathbf{x}_{Nm})\right|.
\]

Общая ошибка складывается из \emph{ошибки дискретизации} численной
схемы и из \emph{статистической ошибки}. Для проверки ошибки
дискретизации с помощью численных экспериментов необходимо, чтобы
статистическая ошибка была меньше ошибки дискретизации.  На практике
это означает, что число $M$ экспериментов методом Монте-Карло должно
быть велико до такой степени, чтобы обеспечить статистическую
погрешность, меньшую ошибки дискретизации.

Для построения доверительного интервала следует провести $K$ серий
экспериментов методом Монте-Карло по $N$ испытаний в каждом. Для
каждой серии следует вычислить \emph{абсолютную среднюю ошибку}:
\[
\mathrm{MAE} = |\mu| = | \bar{f}(\mathbf{x}_{N}) -
\mathbb{E}[f(\mathbf{x}(T))]|.
\]
Тогда получим последовательность из $K$ величин $|\mu_{k}|$,
$k=1,\ldots,K$. Далее вычисляем эмпирическую дисперсию абсолютной
средней ошибки $\sigma^{2}_{\mu}$ по формуле
\[
\sigma^{2}_{\mu} = \frac{1}{K}\sum\limits^{K}_{k=1}(\mu_{k} -
\mu)^{2},
\]
где
\[
\mu =
\frac{1}{MK}\sum\limits^{K}_{k=1}\sum\limits^{M}_{m=1}f(\mathbf{x}_{Nmk})
- \mathbb{E}[f(\mathbf{x}(T))].
\]
Доверительный интервал вычисляется по следующей формуле:
$[\mu - \Delta \mu, \mu + \Delta \mu]$, где
\[
\Delta \mu = t_{1-\frac{\alpha}{2},K}\sqrt{\frac{\sigma^{2}_{mu}}{K}}
\]
$t_{1-\frac{\alpha}{2},K}$ --- $(1-\frac{1}{\alpha})$-квантиль
распределения Стьюдента с $K$ степенью свободы.
\section{Описание библиотеки \texttt{sde}}

\begin{figure}
                \centering
                \includegraphics[width=0.7\linewidth]{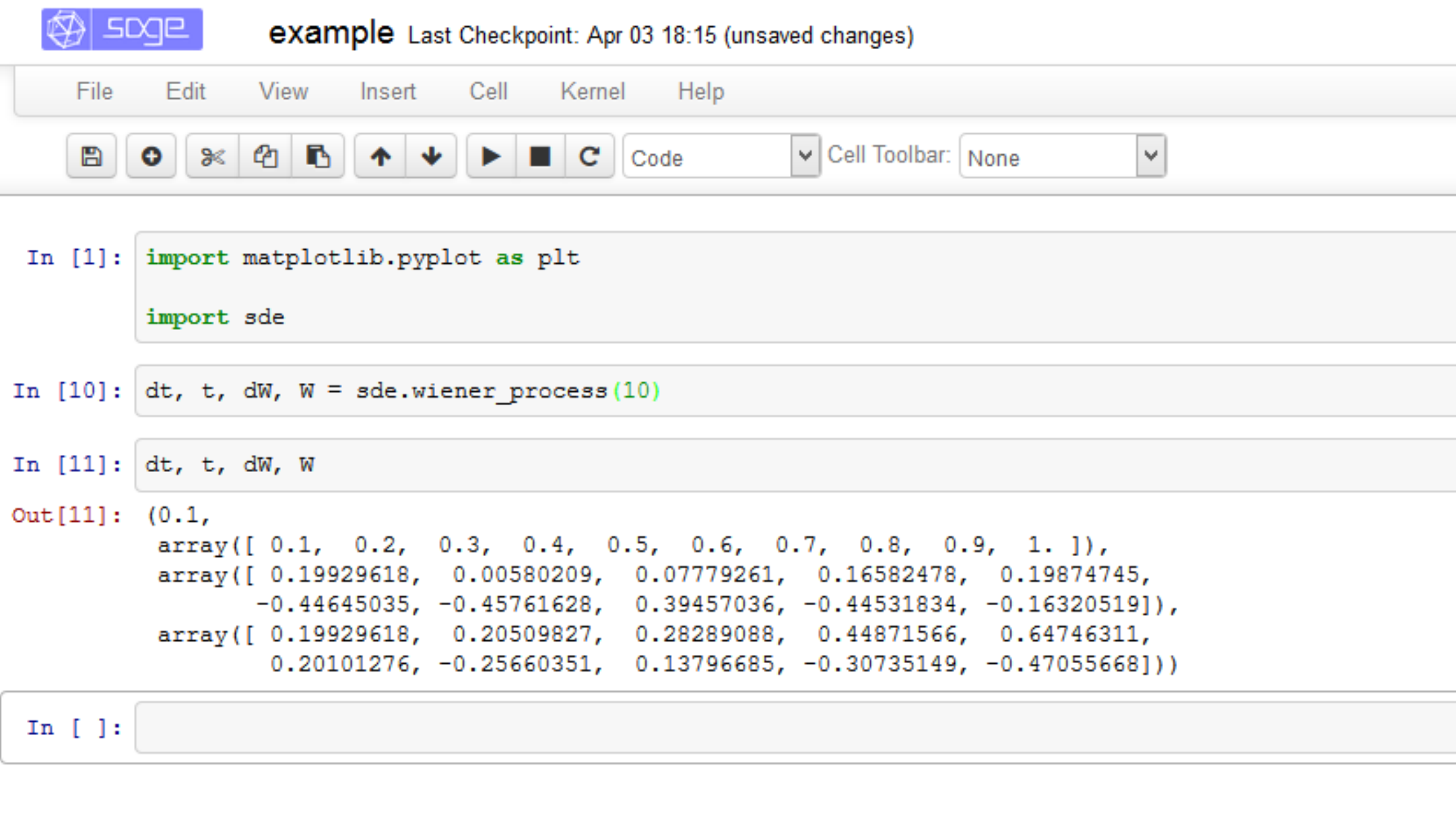}
                \caption{Использование модуля sde в Sage в режиме \texttt{ipython notebook}}
                \label{fig:sage}
        \end{figure}

Для использования библиотеки следует ее подключить как обычный
python-модуль командой \spverb|import sde|. На рис.~\ref{fig:sage}
показан Sage блокнот в режиме \spverb|ipython| (опция
\texttt{--notebook=ipython} при запуске).

В библиотеке содержится ряд функций для внутреннего
использования. Названия таких функций начинаются с двойного нижнего
подчеркивания, как того требуют стилевые правила для кода python PEP8.
\begin{itemize}
\item \spverb|(dt, t) = __time(N, interval=(0.0, 1.0))| --- внутренняя
  функция, которая разбивает временной интервал \texttt{interval} на
  \texttt{N} частей и возвращает \texttt{numpy} массив \texttt{t} с
  шагом \texttt{dt};

\item \spverb|(dW, W) = __scalar_wiener_process(N, dt, seed=None)| ---
  функция генерирует траекторию скалярного винеровского процесса
  \texttt{W} из \texttt{N} частей с шагом \texttt{dt};

\item
  \spverb|(dW, W) = __multidimensional_wiener_process(N, dim, dt, seed=None)|
  --- аналогичная функция для генерации многомерного (размерностью
  \texttt{dim}) процесса Винера;

\item
  \spverb|(dW, W) = __cov_multidimensional_wiener_process(N, dim, dt, seed=None)|
  --- функция генерирует многомерный (размерность \texttt{dim})
  винеровский процесс с использованием функции
  \texttt{numpy.multivariate\_normal}. Процессы $W^1,W^2,\ldots,W^m$
  могут коррелировать, если внутри функции задать матрицу ковариации
  \texttt{cov} не диагональной;

\item
  \spverb|(dt, t, dW, W) = wiener_process(N, dim=1, interval=(0.0, 1.0), seed=None)|
  --- основная функция, которую следует использовать для генерирования
  винеровского процесса. Обязательный аргумент $N$ --- число точек
  разбиения временного интервала \texttt{interval} (по умолчанию
  $[0,1]$). \texttt{dim} задает размерность винеровского
  процесса. Функция возвращает кортеж из четырех элементов, где
  \texttt{dt} --- шаг разбиения ($\Delta t_{i} = h = \mathrm{const}$),
  \texttt{t} --- одномерный \texttt{numpy} массив, содержащий значения
  моментов времени $t_{1},t_{2},\ldots,t_{N}$; \texttt{dW}, \texttt{W}
  --- также \texttt{numpy} массивы размерностью $N\times m$,
  содержащие приращения $\Delta W^{\alpha}_{i}$ и точки выборочной
  траектории $W^{\alpha}_{i}$, где $i=1,\ldots,N;\;\alpha=1,\ldots,m$.
        
\item
  \spverb|__strong_method_selector(name) __weak_method_selector(name)|
  --- заполнение массивов таблицы Бутчера для конкретного метода со
  слабой и с сильной сходимостью.
\end{itemize}

Следующий ряд функций служит для аппроксимирования интегралов Ито для
методов с сильной сходимостью. Все они принимают в качестве
обязательного аргумента массив приращений винеровского процесса
\texttt{dW}, шаг разбиения временного интервала \texttt{h} и
возвращают список значений интегралов на каждом шаге траектории. Для
кратных интегралов возвращается список массивов:
\begin{itemize}
\item \spverb|Ito1W1(dW)| --- генерирование значений однократного
  интеграла Ито для скалярного винеровского процесса;
\item \spverb|Ito2W1(dW, h)| --- генерирование значений двукратного
  интеграла Ито для скалярного винеровского процесса;
\item \spverb|Ito1Wm(dW, h)| --- генерирование значений однократного
  интеграла Ито для многомерного винеровского процесса;
\item \spverb|Ito2Wm(dW, h, n)| --- аппроксимация значений двукратных
  интегралов Ито. Возвращает список массивов с элементами $I^{ij}$ для
  каждого шага траектории. Аргумент \texttt{n} отвечает за число
  слагаемых в бесконечном ряде $\mathbf{A}(h)$;
\item \spverb|Ito3W1(dW, h)| --- генерирование трехкратного интеграла
  Ито для скалярного винеровского процесса.
\end{itemize}

Теперь перейдем к основным функциям, реализующим численные методы с
сильной сходимостью:
\begin{itemize}
\item
  \spverb|EulerMaruyama(f, g, h, x_0, dW) EulerMaruyamaWm(f, g, h, x_0, dW)|
  --- метод Эйлера--Маруямы для скалярного и многомерного винеровских
  процессов, где \texttt{f} --- вектор сноса, \texttt{g} --- матрица
  диффузии;
        
\item \spverb|strongSRKW1(f, g, h, x_0, dW, name='SRK2W1')| ---
  функция для решения СДУ со скалярным винеровским процессом, где
  \texttt{f(x)} и \texttt{g(x)} те же, что и в предыдущем методе;
  \texttt{h} --- шаг сетки, \texttt{x\_0} --- начальное значение,
  \texttt{dW} --- массив приращений винеровского процесса \texttt{N};
  аргумент \texttt{name} может принимать значения \texttt{'SRK1W1'} и
  \texttt{'SRK2W2'};

\item \spverb|oldstrongSRKp1Wm(f, G, h, x_0, dW, name='SRK1Wm')| ---
  стохастический метод Рунге--Кутты сильного порядка $p = 1.0$ для
  многомерного винеровского процесса. Функция реализует алгоритм с
  использованием вложенных циклов, поэтому работает крайне медленно и
  оставлена лишь для тестирования;

\item \spverb|strongSRKp1Wm(f, G, h, x_0, dW, name='SRK1Wm')| ---
  стохастический метод Рунге--Кутты сильного порядка $p = 1.0$ для
  многомерного винеровского процесса. Используется метод
  \texttt{numpy.einsum}, о котором говорилось выше. Благодаря этому
  методу вычисление свертки четырехмерного массива значительно
  ускоряется.
\end{itemize}

Следующая группа функций реализует методы со слабой сходимостью:
\begin{itemize}
\item \spverb|n_point_distribution(values, probabilities, shape)| ---
  многоточечное распределение, в котором \spverb|values| --- массив
  принимаемых случайной величиной значений, \spverb|probabilities| ---
  массив вероятностей выпадения этих значений, \spverb|shape| --- форма
  этих массивов;
        
\item \spverb|weakIto(h, dim, N)| --- замена интегралов Ито в случае
  слабой сходимости;

\item \spverb|weakSRKp2Wm(f, G, h, x_0, dW, name='SRK1Wm')| ---
  стохастический метод Рунге--Кутты слабого порядка $p = 2.0$ для
  многомерного винеровского процесса, где массив \texttt{dW}
  используется только для вычисления размерностей и количества точек в
  траектории, а сам алгоритм метода для вычислений его не использует.
\end{itemize}

Как уже упоминалось выше, для использования стохастических численных
методов со слабой сходимостью следует применять метод Монте--Карло,
который заключается в многократном применении одного и того же метода
к СДУ. Это очень ресурсоемкий процесс, который, однако, легко
распараллеливается как на многопроцессорном компьютере с общей
памятью, так и на кластере с несколькими узлами. Для этой цели можно
использовать модуль \spverb|multiprocessing| из стандартной библиотеки
языка python. Этот модуль позволяет автоматически создать необходимое
число процессов и распределить работу между ними. Так как каждое
вычисление производится полностью независимо, то процессам не нужно
обмениваться между собой никакими данными.

Важно заметить, что \spverb|multiprocessing| для создания процессов в
рамках операционных систем Unix использует системный вызов
\spverb|fork|, из-за чего все порожденные процессы наследуют один и тот
же генератор случайных чисел, в результате чего генерируют совершенно
одинаковые винеровские траектории. Во избежание этого во всех функциях
библиотеки, которые так или иначе связаны с генерацией случайных
чисел, предусмотрен необязательный аргумент \spverb|seed|, который может
принимать целочисленные значения и использоваться для инициализации
генератора. При параллельных вычислениях в качестве аргумента
\spverb|seed| можно передать порядковый номер вычисления. Более
подробное описание доступно по ссылке
\url{https://bitbucket.org/mngev/sde-numerical-integrators}.

\section{Стохастическая модель хищник-жертва}

Описанный выше комплекс программ был применен для численных
экспериментов со стахостичекой моделью
хищник-жертва~\cite{kulyabov:2013:conf:mmcp,kulyabov:2014:icumt-2014:p2p,ef-kor-gev-kul-sev:vestnik-miph:2014-3}.

Модель задается СДУ с вектором сноса $f$ и матрицей диффузии $G$
следующего вида:

\begin{equation}
f(x,y) = \left(
  \begin{array}{c}
    k_{1}ax - k_{2}xy\\
    k_{2}xy - k_{3}y
  \end{array}
\right) \qquad G(x,y) = \left(
  \begin{array}{cc}
    k_{1}ax+k_{2}xy & -k_{2}xy\\
    -k_{2}xy & k_{2}xy + k_{3}y
  \end{array}
\right)^{\frac{1}{2}},
\end{equation}
где $x$ --- численность жертв, $y$ --- численность хищников, $a$ ---
рост жертв, $k_{1}$ --- конкуренция жертв, $k_{2}$ --- частота
хищничества, $k_{3}$ --- смертность хищников. Выберем параметры
$a, k_1, k_2, k_3 = (0.5, 3.0, 0.05, 2.5)$ и начальную точку, близкую
к стационарной $(x_0, y_0) = (50.0, 30.0)$.

Для стохастического случая важное значение имеет величина
\[
\mathrm{d}I(x,y) = \frac12 \left( \frac{ak_{1}k_{3}}{x} +
  \frac{k_{2}k_{3}y}{x} + \frac{ak_{1}k_{2}x}{y} +
  \frac{ak_{1}k_{3}}{y} \right),
\]
которая характеризует фазовый объем системы и в среднем возрастает в
течение эволюции системы~\cite{kulyabov:2013:conf:mmcp}. Реализованные
в библиотеке слабые и сильные методы были применены к данной задаче
для проверки свойств модели с помощью численного моделирования.

С помощью слабого метода Рунге--Кутты было вычислено 500 конкретных
реализаций траекторий решения СДУ. На основе этих данных были найдены
различные статистические характеристики, такие как медиана и
процентили. Эволюция этих характеристик с течением времени
иллюстрирует характерные особенности стохастической системы
хищник-жертва. Так из рис.~\ref{fig:prepre4} видно неуклонное
возрастание фазового объема со временем, также как из
рис.~\ref{fig:prepre5}, где изображена эволюция инварианта
$\mathrm{d}I$ со временем. Видно, что $\mathrm{d}I$ в среднем
возрастает, хотя и не монотонно.

\begin{figure}
  \centering
  \includegraphics[width=0.6\linewidth]{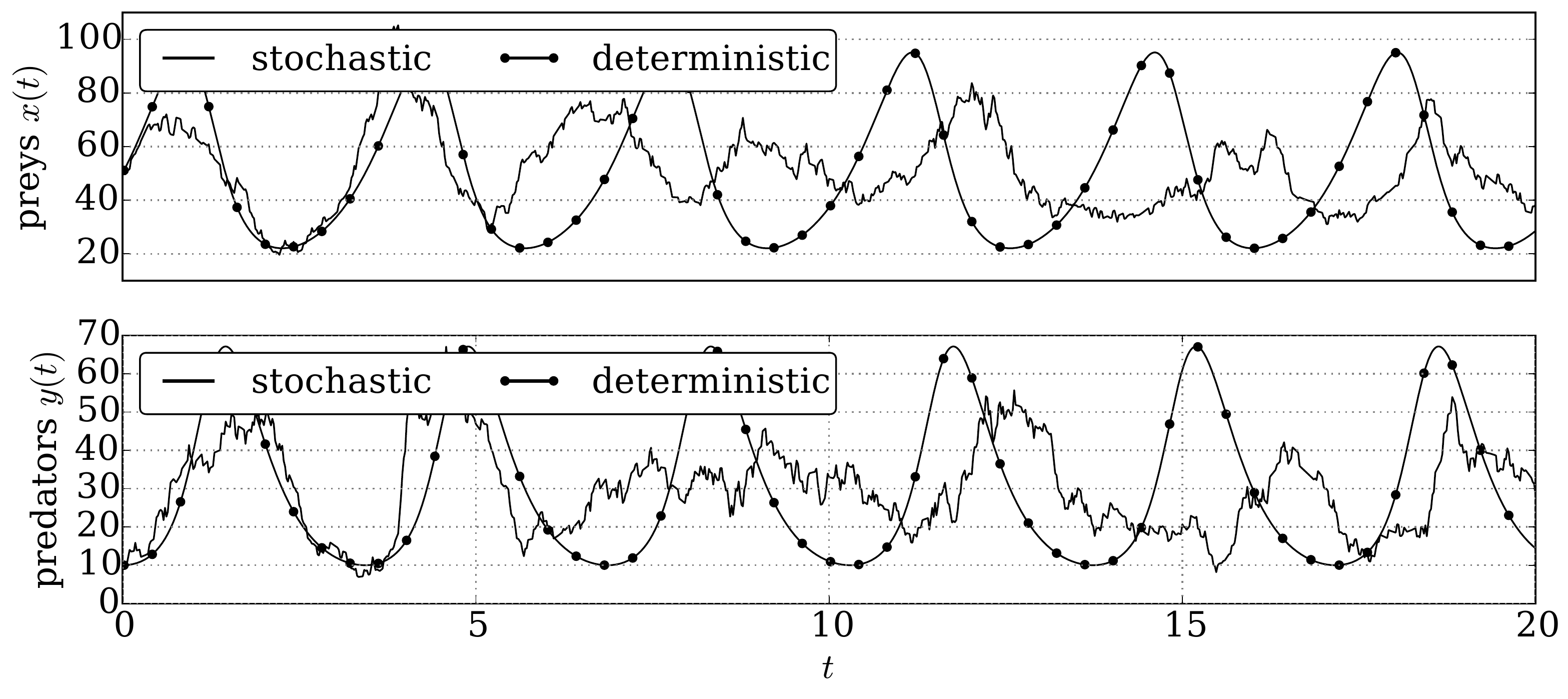}
  \caption{Графики решений уравнений детерминированной модели
    хищник-жертва и соответствующей стохастической модели с помощью
    стохастического метода Рунге--Кутты с сильной сходимостью}
  \label{fig:prepre1}
\end{figure}

\begin{minipage}[b]{0.45\linewidth}
\begin{figure}[H]
  \centering
  \includegraphics[width=1.0\linewidth]{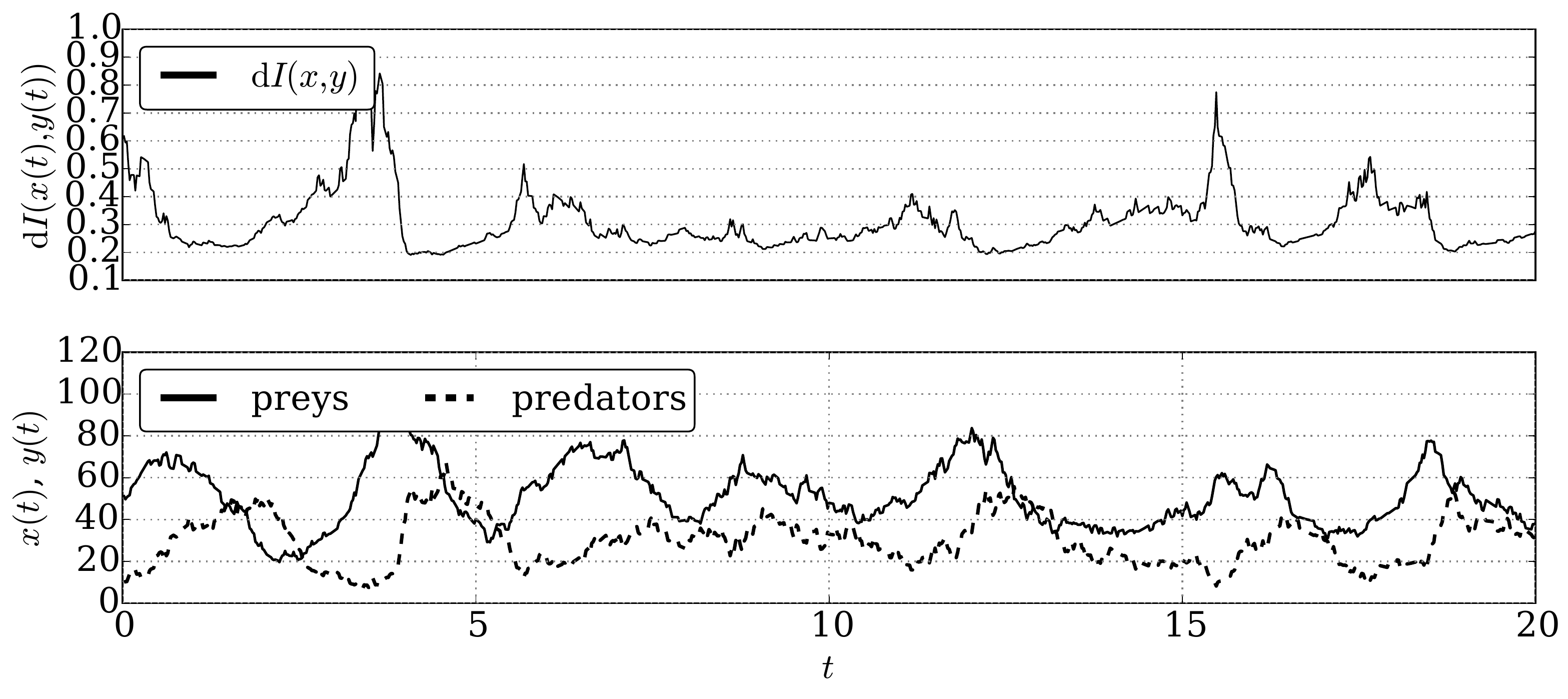}
  \caption{Эволюция инварианта $\mathrm{d}I$ во времени для конкретной
    реализации траектории стохастической модели хищник-жертва,
    полученной методом Рунге--Кутты с сильной сходимостью}
  \label{fig:prepre2}
\end{figure}
\end{minipage}
\hfill
\begin{minipage}[b]{0.45\linewidth}
\begin{figure}[H]
  \centering
  \includegraphics[width=1.0\linewidth]{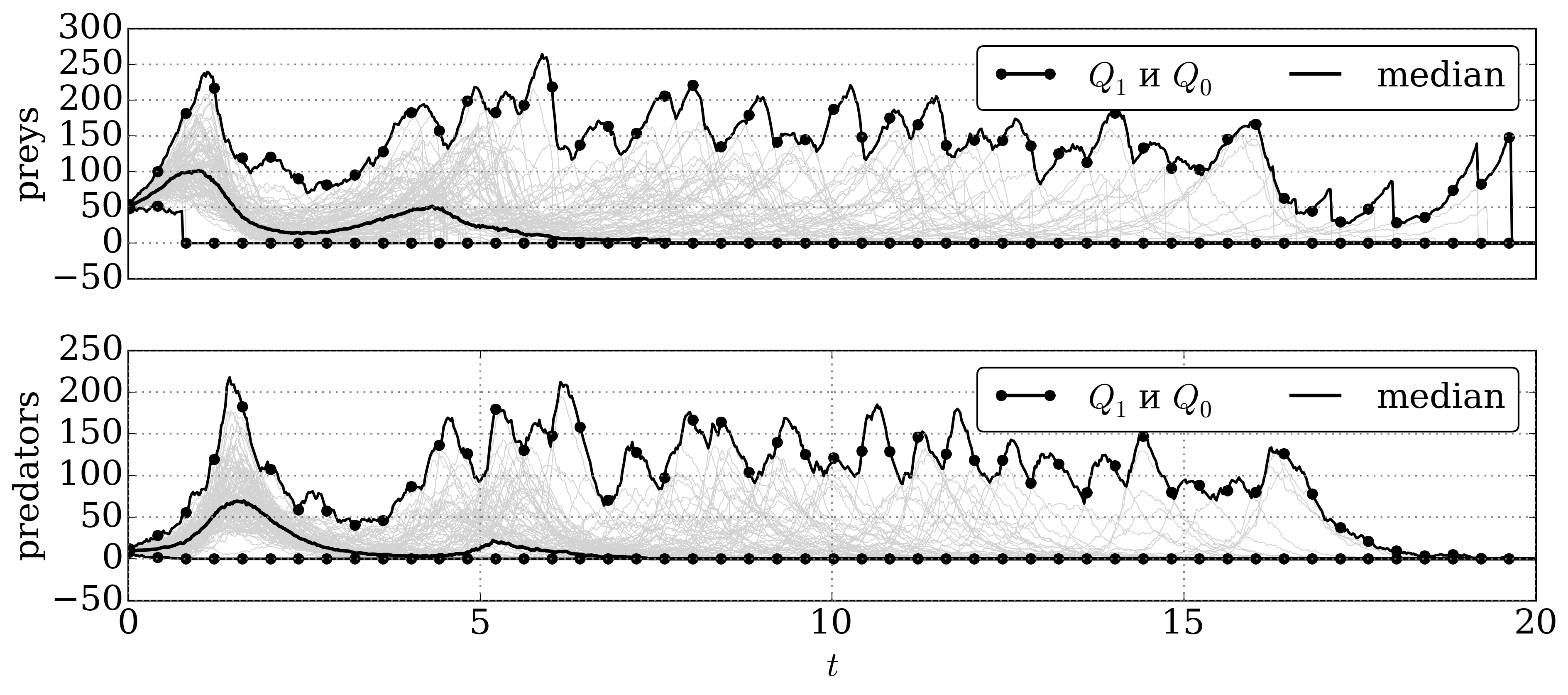}
  \caption{Множество траекторий решений стохастической модели
    хищник-жертва (серый) и соответствующие данному множеству
    траекторий процентиль $Q_{1}, Q_{0}$ и медиана}
  \label{fig:prepre3}
\end{figure}
\end{minipage}

\begin{minipage}[b]{0.45\linewidth}
\begin{figure}[H]
  \centering
  \includegraphics[width=1.0\linewidth]{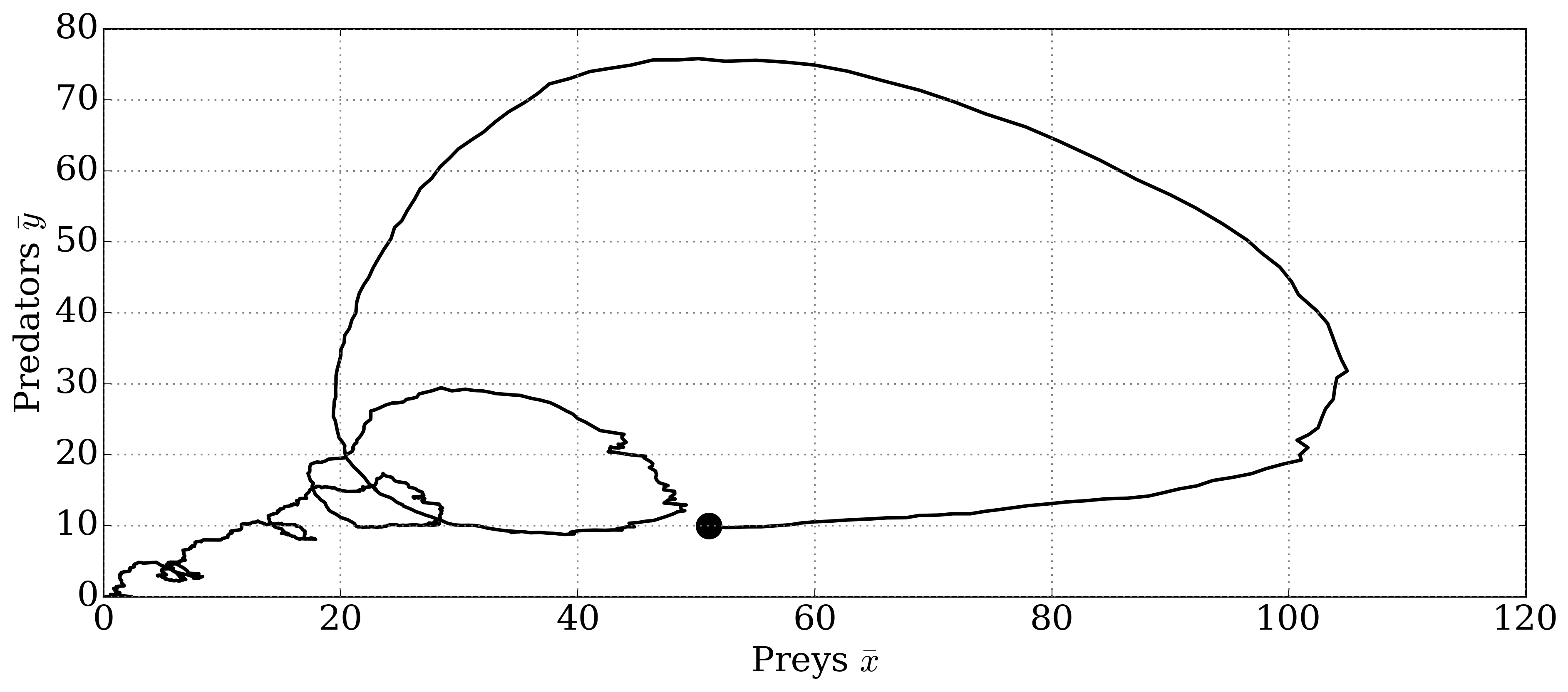}
  \caption{Фазовый портрет выборочных средних от численности хищников
    и жертв на основании 500 траекторий, полученных методом
    Рунге--Кутты со слабой сходимостью}
  \label{fig:prepre4}
\end{figure}
\end{minipage}
\hfill
\begin{minipage}[b]{0.45\linewidth}
\begin{figure}[H]
  \centering
  \includegraphics[width=1.0\linewidth]{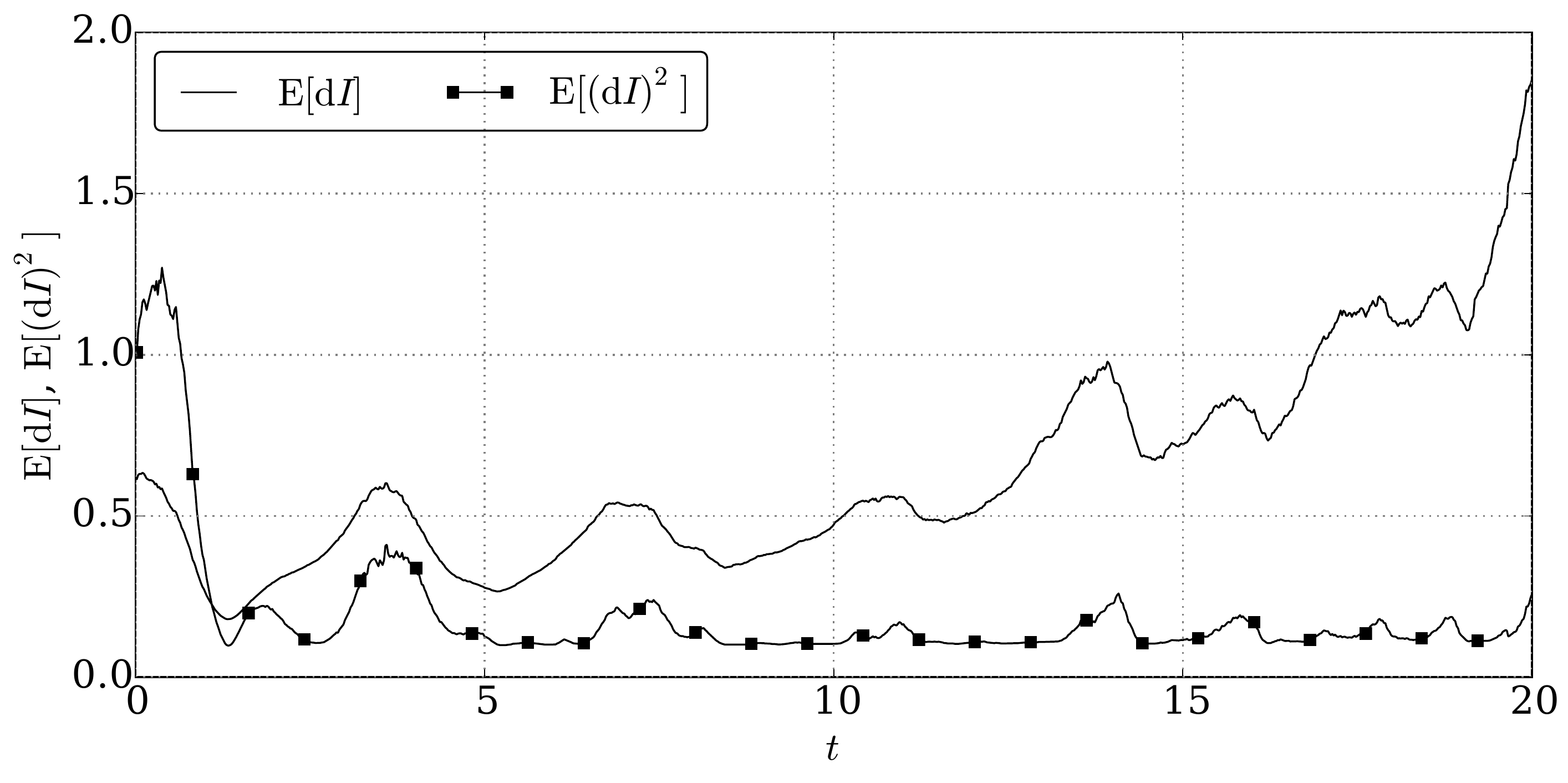}
  \caption{Первый и второй эмпирические моменты инварианта
    $\mathrm{d}I$ на основании 500 траекторий, полученных методом
    Рунге--Кутты со слабой сходимостью}
  \label{fig:prepre5}
\end{figure}
\end{minipage}

  Другое характерное отличие стохастической от классической
  детерминированной модели проявляется в тенденции к неизбежной гибели
  популяций в процессе эволюции. На рис.~\ref{fig:prepre3} серым
  цветом обозначены все полученные траектории. Видно, что по мере
  эволюции системы кривые траекторий пересекают ось абсцисс, что
  означает гибель одного из представителя популяций. Эта особенность
  стохастической модели выгодно отличает ее от детерминированной, в
  которой гибель одного из видов не возможна в принципе.

\section{Заключение}
\label{sec:conclusion}

Мы рассмотрели несколько конкретных реализаций стохастического метода
Рунге--Кутты. На данный момент в библиотеке используются методы,
найденные А.~Росслером~\cite{Rossler_2010} как наиболее эффективные из
известных на данный момент авторам. Однако основные функции
\texttt{strongSRKW1} и \texttt{weakSRKp2Wm} написаны в соответствии с
общим алгоритмом и могут использовать любую таблицу Бутчера
соответствующей стадийности. Это позволяет пользователю библиотеки
расширить ее функционал, добавив новые методы.

Исходный код библиотеки доступен по адресу
\url{https://bitbucket.org/mngev/sde-numerical-integrators}. Там же
находятся блокноты со всеми вышеизложенными примерами применения
библиотечных функций для решения СДУ. Авторы предполагают активно
развивать и дорабатывать библиотеку, добавляя в нее новые функции.

\def\acknowledgmentsname{Благодарности}

\begin{acknowledgments}

Работа частично поддержана грантами РФФИ \textnumero 14-01-00628,
15-07-08795 и 16-07-00556. 

Расчёты проведены на вычислительном кластере «Felix» РУДН и на
Гетерогенном вычислительном кластере «HybriLIT» Многофункционального
центра хранения, обработки и анализа данных ОИЯИ.

Опубликовано в:
\emph{Gevorkyan~M.~N., Velieva~T.~R., Korolkova~A.~V., Kulyabov~D.~S.,
  Sevastyanov~L.~A.}
Stochastic Runge–Kutta Software Package for Stochastic Differential Equations~//
Dependability Engineering and Complex Systems.~---
\newblock Springer International Publishing, 2016.~---
\newblock Vol.~470.~---
\newblock P.~169--179.~---
\href{http://dx.doi.org/10.1007/978-3-319-39639-2_15}{doi:10.1007/978-3-319-39639-2\_15}.

Исходные тексты:
\url{https://bitbucket.org/yamadharma/articles-2015-rk-stochastic}

\end{acknowledgments}

\bibliographystyle{gost2008l}

\bibliography{bib/rk-stoch-realis/bib,bib/rk-stoch-realis/self}